\documentclass{aa}  
\usepackage{graphicx}
\usepackage{placeins}
\usepackage{multirow}
\usepackage{natbib}
\bibpunct{(}{)}{;}{a}{}{,}
\usepackage{upgreek}
\usepackage{amsmath}
\usepackage{wasysym}
\usepackage{amssymb}
\usepackage{txfonts}
\usepackage[linkcolor=blue, bookmarks=true, bookmarksopen=true,
citecolor=blue, urlcolor=blue, colorlinks=true,
breaklinks=true, plainpages=false]{hyperref}
\usepackage[all]{hypcap}
\newcommand{\mdegr}{^{\circ}}
\newcommand{\sub}[1]{_\mathrm{#1}}
\newcommand{\dif}{\mathrm{d}}
\defcitealias{2014A&A...572A..76V}{Paper~I}

\graphicspath{{./img/}}

\begin{document}

\title{Dusty tails of evaporating exoplanets}
\subtitle{II. Physical modelling of the KIC~12557548b light curve}
\titlerunning{Dusty tails of evaporating exoplanets. II.}

  \author{
    R. van Lieshout\inst{1,2}
    \and
    M. Min\inst{3,1}
    \and
    C. Dominik\inst{1,4}
    \and
    M. Brogi\inst{5,6,}\thanks{NASA Hubble Fellow}
    \and
    T. de Graaff\inst{1}
    \and
    S. Hekker\inst{7,8}
    \and
    M. Kama\inst{6,1}
    \and
    \mbox{C. U. Keller\inst{6}}
    \and
    \mbox{A. Ridden-Harper\inst{6}}
    \and
    \mbox{T. I. M. van Werkhoven\inst{6}}
    }

  \institute{
    Anton Pannekoek Institute for Astronomy, University of Amsterdam, Science Park 904, 1098 XH Amsterdam, The Netherlands
    \\ \email{lieshout@ast.cam.ac.uk}
    \and
    Institute of Astronomy, University of Cambridge, Madingley Road, Cambridge CB3 0HA, UK
    \and
    SRON Netherlands Institute for Space Research, Sorbonnelaan 2, 3584 CA Utrecht, The Netherlands
    \and
    Department of Astrophysics/IMAPP, Radboud University Nijmegen, P.O. Box 9010, 6500 GL Nijmegen, The Netherlands
    \and
    CASA, Department of Astrophysical and Planetary Sciences, University of Colorado, 389-UCB, Boulder, CO 80309, USA
    \and
    Leiden Observatory, Leiden University, P.O. Box 9513, 2300 RA Leiden, The Netherlands
    \and
    Max Planck Institute for Solar System Research, Justus-von-Liebig-Weg 3, D-37077 G\"{o}ttingen, Germany
    \and
    Stellar Astrophysics Centre, Department of Physics and Astronomy, Aarhus University, Ny Munkegade 120, DK-8000 Aarhus C, Denmark
    }


 
  \abstract
   {
   Evaporating rocky exoplanets,
   such as KIC~12557548b,
   eject large amounts of dust grains,
   which can trail the planet in a comet-like tail.
   When such objects occult their host star,
   the resulting transit signal
   contains information about the dust in the tail.
   }
   {
   We aim to use the detailed shape of the \textit{Kepler} light curve of KIC~12557548b
   to constrain the size and composition of the dust grains that make up the tail,
   as well as the mass loss rate of the planet.
   }
   {
   Using a self-consistent numerical model of the dust dynamics and sublimation,
   we calculate the shape of the tail
   by following dust grains from their ejection from the planet
   to their destruction due to sublimation.
   From this dust cloud shape, we generate synthetic light curves
   (incorporating the effects of extinction and angle-dependent scattering),
   which are then compared with the phase-folded \textit{Kepler} light curve.
   We explore the free-parameter space thoroughly using a Markov chain Monte Carlo method.
   }
   {
   Our physics-based model is capable of reproducing the observed light curve in detail.
   Good fits are found for initial grain sizes between 0.2 and 5.6~$\upmu$m
   and dust mass loss rates of 0.6 to 15.6~M$\sub{\oplus}$~Gyr$^{-1}$ ($ 2 \sigma $ ranges).
   We find that only certain combinations of material parameters yield the correct tail length.
   These constraints are consistent with dust made of corundum (Al$_2$O$_3$),
   but do not agree with a range of carbonaceous, silicate, or iron compositions.
   }
   {
   Using a detailed, physically motivated model,
   it is possible to constrain the composition of the dust in the tails of evaporating rocky exoplanets.
   This provides a unique opportunity to probe to interior composition of the smallest known exoplanets.
   }

  \keywords{ eclipses -- occultations -- planets and satellites: composition
    -- planets and satellites: individual: KIC~12557548b -- planet-star interactions }

   \maketitle
%

\section{Introduction}

Determining the chemical composition of exoplanets
is an important step in advancing our understanding of the Earth's galactic neighbourhood
and provides valuable benchmarks for theories of planet formation and evolution.
For small (i.e., Earth-sized and smaller) exoplanets,
most efforts so far have been directed at
determining a planet's mean density from independent measurements of its size and mass,
which gives an indication of the bulk composition \citep{2007ApJ...659.1661F,2013Natur.503..381H}.
This method, however, is restricted by the observational lower limits for which planetary radii and masses can reliably be determined
\citep[although recent progress is pushing the limits ever further down;][]{2015Natur.522..321J}.
Another, more fundamental, problem is that
exoplanets with different (combinations of) chemical compositions can have the same mean density,
making it impossible to distinguish between these compositions using just the planet's size and mass \citep{2007ApJ...669.1279S}.
In particular, whether small carbon-based exoplanets exist
is an open question that cannot be resolved with bulk density measurements alone \citep[see Fig.~9 of][]{2007ApJ...669.1279S,2012ApJ...759L..40M}.

Another method of investigating the chemical composition of exoplanetary material
is the study of white-dwarf atmospheres that are polluted by the accretion of tidally disrupted asteroids or minor planets
(see \citealt{2014AREPS..42...45J} for a review
and \citealt{2015Natur.526..546V}; \citealt{2016ApJ...816L..22X} for some recent results).
This method allows the bulk composition of the accreted bodies to be measured with unprecedented precision.
However, it can only be applied to white-dwarf systems,
and the exact relation between the measured compositions and those of the exoplanets in the original, main-sequence-stage planetary systems is not yet clear.

\begin{figure*}[!t]
  \includegraphics[width=\linewidth]{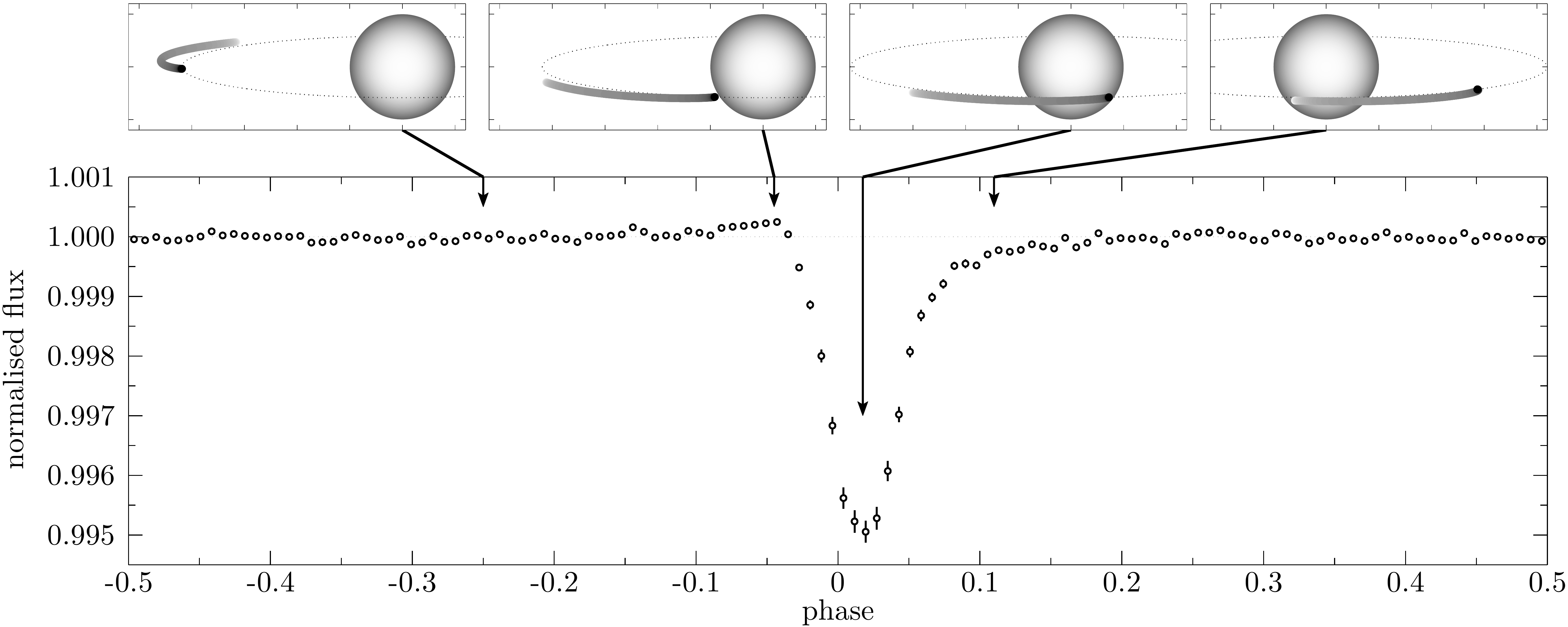}
  \caption{
  The phase-folded long-cadence \textit{Kepler} light curve of KIC~1255b (\textbf{bottom}),
  together with schematic views of the system at different orbital phases (\textbf{top}),
  illustrating how an asymmetric dust cloud can explain the peculiar transit profile.
  Arrows indicate which sketch corresponds to which orbital phase.
  For details on the observational data, see Sect.~\ref{s:meth_obs} of this work and Sect.~2 of \citet{2014A&A...561A...3V}.
  The error bars on flux include the spread caused by the variability in transit depth,
  making the in-transit error bars greater than the out-of-transit ones
  (which are mostly smaller than the size of the symbols).
  In the sketches, the star, the orbit of the planet, and the length of the dust tail are all drawn to scale;
  the tick marks on the axes are spaced one stellar radius apart.
  The vertical thickness of the dust cloud and its colour gradient
  (which illustrates the gradually decreasing dust density)
  are chosen for illustrative purposes.
  }
  \label{fig:intro_sketch}
\end{figure*}

The discovery of transiting evaporating rocky exoplanets \citep{2012ApJ...752....1R}
has opened up a possible new channel for determining the chemical compositions of small exoplanets
that is complementary to the methods mentioned above.
Through the evaporation of their surface,
these objects present material from their interior to the outside,
where it can be examined as it blocks and scatters star light.
We recently showed how the composition of the outflowing material
can be determined from the shape of the object's transit light curve
using semi-analytical expressions
\citep[][hereafter \citetalias{2014A&A...572A..76V}]{2014A&A...572A..76V}.
In the present paper, we revisit this problem using a numerical model,
which allows us to let go of several of the simplifying assumptions made in \citetalias{2014A&A...572A..76V}
and to use more directly all the information contained in the light curve.

\subsection{Evaporating rocky exoplanets}

To date, there are three known (candidates of) transiting evaporating rocky exoplanets:
KIC~12557548b \citep[hereafter KIC~1255b;][]{2012ApJ...752....1R},
\mbox{KOI-2700b} \citep{2014ApJ...784...40R},
and \mbox{K2-22b} \citep{2015ApJ...812..112S},
all three discovered using the \textit{Kepler} telescope \citep{2010Sci...327..977B}.\footnote{%
A search for more such objects amongst short-period \textit{Kepler} exoplanet candidates
did not find any additional ones \citep{2014AN....335.1018G}.}
They all orbit K- and \mbox{M-type} main-sequence stars
in orbital periods of less than a day
and their light curves are marked by asymmetric transit profiles and variable transit depths.
Both light-curve properties can be explained by a scenario in which the extinction of star light is caused by
an asymmetric cloud of dust grains,
whose collective cross-section changes from transit to transit.
This scenario was first put forward by \citet{2012ApJ...752....1R} for the prototype KIC~1255b; we briefly summarise it here.

The dust grains that make up the cloud originate in a small evaporating planet.
Once they have left the planet,
radiation pressure from the host star pushes them
into a comet-like tail trailing the planet.
With increasing distance from the planet, the dust grains speed up with respect to the planet.
They also gradually sublimate due to the intense stellar irradiation, decreasing their size.
Both effects cause the angular density of extinction cross-section to decrease further into the tail.
The resulting asymmetric shape of the dust cloud
can explain the sharp ingress and gradual egress of the observed transit light curve (see Fig.~\ref{fig:intro_sketch}).\footnote{%
\label{fn:k2_22b}
The light curve of \mbox{K2-22b} has a markedly different shape, 
which can be explained by a streamer of dust grains leading the planet (instead of trailing it).
In this object, whose host star is less luminous, the initial launch velocities of the dust grains
(rather than radiation pressure) may dominate the dynamics \citep{2015ApJ...812..112S}.
}
In addition, scattering of star light by dust grains results in a brightening just before the transit,
when the bulk of the dust cloud is not in front of (the brightest part of) the stellar disk,
but close enough to yield small scattering angles.
The asymmetry of the dust cloud means that this effect is much stronger around ingress than egress.

The dust cloud scenario has been validated by the colour dependence of the transit depth \citep{2015ApJ...800L..21B,2015ApJ...812..112S,2016arXiv160507603S},
while many false positive scenarios for this type of event have been ruled out
based on radial velocity measurements, high angular resolution imaging, and photometry \citep{2014ApJ...786..100C}.
Morphological modelling of the KIC~1255b light curve has allowed some properties of its dust cloud to be determined
\citep{2012A&A...545L...5B,2013A&A...557A..72B,2014A&A...561A...3V}.
In particular, both the wavelength dependence of the transit depth and the morphological dust cloud models
indicate that the dust grains have radii in the range 0.1 to 1.0~$\upmu$m.

To explain the variation in transit depth,
the dust cloud scenario invokes
erratic variations in the planet's dust production rate.
By making some assumptions about the dust grains,
it is also possible to infer the average dust mass loss rate of the evaporating planet
(i.e., excluding the mass lost in gas) from the light curve.
For KIC~1255b and \mbox{K2-22b}, the dust mass loss rates are estimated to be of the order of 0.1 to 1~M$\sub{\oplus}$~Gyr$^{-1}$
(\citealt{2012ApJ...752....1R,2013MNRAS.433.2294P,2013ApJ...776L...6K}; \citetalias{2014A&A...572A..76V}; \citealt{2015ApJ...812..112S}),
while for \mbox{KOI-2700b} it may be one to two orders of magnitude lower (\citealt{2014ApJ...784...40R}; \citetalias{2014A&A...572A..76V}).

The planet's mass loss is thought to be fuelled by the total bolometric flux from the host star \citep{2012ApJ...752....1R}.\footnote{%
\mbox{X-ray}-and-ultraviolet-driven evaporation was suggested as an alternative, based on a relation between transit depth and stellar rotational phase \citep{2013ApJ...776L...6K}.
A more straightforward explanation for this relation, however, is the occultation of starspots by the transiting dust cloud \citep{2015MNRAS.449.1408C}.
}
Stellar radiation heats the planetary surface to
a temperature exceeding 2000~K,
which causes the solid surface to vaporise,
creating a metal-rich atmosphere
\citep[as has been modelled in detail for super-Earths;][]{2009ApJ...703L.113S,2011ApJ...742L..19M,2012ApJ...755...41S}.
This atmosphere is hot and expands into the open space around the planet,
driving a ``Parker-type'' thermal wind \citep{2012ApJ...752....1R}.
As the gas expands and cools, its refractory constituents can condense into dust grains.\footnote{%
Another mechanism that could be responsible for loading the planet's atmosphere with dust is explosive volcanism \citep{2012ApJ...752....1R}.}
Small dust grains are entrained in the gas flow until the gas thins out,
from which point the dust dynamics are controlled by stellar gravity and radiation pressure.

\citet{2013MNRAS.433.2294P} modelled the planetary outflow in detail,
finding that the mass loss rate is a strong function of the mass of the evaporating body.
According to their model, the mass loss rate of KIC~1255b indicates that
the planet cannot be more massive than about 0.02~M$\sub{\oplus}$ (i.e., less than twice the mass of the Moon)
and will disintegrate completely within about 40 to 400~Myr.
The planetary radius corresponding to the mass limit
is consistent with the upper limits on the size of the planet --
derived from the non-detection of transits in some parts of the light curve \citep{2012A&A...545L...5B}
and secondary eclipses in the entire light curve \citep{2014A&A...561A...3V} --
and, if correct, would make KIC~1255b one of the smallest exoplanets known.

\subsection{Dusty tail composition}

Regardless of how exactly the evaporating planet produces and ejects dust,
the composition of the dust in the tail will reflect that of the planet.
The precise relation between the two compositions may be complicated by selection effects
such as preferential condensation of certain dust species in the atmosphere \citep[e.g., Sect.~3.2.2 of][]{2012ApJ...755...41S}
and possibly the fractional vaporisation of a magma ocean \citep[Sect.~6.1 of][]{2011Icar..213....1L}.
Nevertheless, identifying the composition of the dust can lead to insights into
the composition of the surface of the planet and possibly its interior
(if prior evaporation has already removed the original surface, exposing deeper layers).
Such insights are invaluable for theories of planet formation and evolution.

Building upon the work of \citet{2002Icar..159..529K} and \citet{2012ApJ...752....1R,2014ApJ...784...40R},
we recently demonstrated how the length of a dusty tail trailing an evaporating exoplanet
can be used to constrain the composition of the dust in this tail \citepalias{2014A&A...572A..76V}.
In a nutshell, the length of the tail is determined by the interplay of radiation-pressure-induced azimuthal drift of dust grains
and the decrease in size of these grains due to sublimation.
Because the sublimation rate of the dust is strongly dependent on its compositions,
the tail length is a proxy for grain composition.
By comparing tail-length predictions for potential dust species
with the observed tail length (derived from the duration of the transit egress),
it is possible to put constraints on the composition of the dust in the tail.

\citetalias{2014A&A...572A..76V} presents a semi-analytical description of dusty tails, in which
the shape of the tail is described using just two parameters:
the tail's characteristic length and its initial angular density.
The values of these two parameters are taken from the morphological tail models,
which derive them from light curve fitting \citep{2012A&A...545L...5B,2013A&A...557A..72B,2014A&A...561A...3V,2014ApJ...784...40R}.
However, describing the tail morphology in just two parameters ignores many details of the tail's shape,
which may be used to constrain the dust composition from the detailed shape of the transit light curve.
Furthermore, the derivation of a semi-analytical description of the dust tail in \citetalias{2014A&A...572A..76V} requires many assumptions,
which may undermine the applicability of the resulting equations.

To take the next step in modelling the dusty tails of evaporating planets,
it is desirable to employ a physics-based (in contrast to morphological) model of the tail
that self-consistently takes into account the interplay of
grain-size-dependent radiation-pressure dynamics and temperature-dependent grain-size evolution.
In this paper, we develop such a model.
In brief, the model consists of a particle-dynamics-and-sublimation simulation,
followed by a transit-profile generation using a light-scattering code.
Similar modelling work has been done previously to predict the light curves due to possible extrasolar comets in the $\upbeta$~Pictoris system
\citep{1999A&A...343..916L,1999A&AS..140...15L},
with the major differences that
these comets have orbital periods of years rather than hours
and sublimation does not have to be taken into account.
For the dusty tails of evaporating planets,
\citet[][their Sect.~4.6]{2012ApJ...752....1R}
and \citet[][their Sect.~6.2]{2015ApJ...812..112S}
did particle-dynamics simulations,
but using a constant lifetime of the dust grains against sublimation
and without generating light curves.
In order to derive constraints on the dust composition from broadband transit profiles,
it is essential to treat dust sublimation in a self-consistent, time-dependent way.

\begin{table}
  \centering
  \small
  \caption{Host star and system parameters of KIC~1255b}
  \label{tbl:sys_pars}
  {
  \renewcommand{\arraystretch}{1.16}
  \begin{tabular}{lccc}
  \hline
  \hline
  Parameter & Symbol & Units & Value \\
  \hline
  Stellar effective temperature & $ T\sub{eff,\star} $ & K & $ 4550\substack{+140 \\ -131} $ \\
  Stellar mass & $ M\sub{\star} $ & M$ \sub{\odot} $ & $ 0.666\substack{+0.067 \\ -0.059} $ \\
  Stellar radius & $ R\sub{\star} $ & R$ \sub{\odot} $ & $ 0.660\substack{+0.060 \\ -0.059} $ \\
  Stellar luminosity & $ L\sub{\star} $ & L$ \sub{\odot} $ & $ 0.168\substack{+0.037 \\ -0.036} $ \\
  \hline
  Planet's orbital period & $ P\sub{p} $ & days & $ 0.6535538(1) $ \\
  Planet's semi-major axis & $ a\sub{p} $ & AU & $ 0.0129(4) $ \\
  \hline
  \end{tabular}
  \tablefoot{
  The stellar parameters are taken from \citet{2014ApJS..211....2H};
  the planet's orbital period is from \citet{2014A&A...561A...3V}.
  Numbers in brackets indicate the uncertainty on the last digit.
  }
  }
\end{table}

We apply our model to the prototypical evaporating rocky exoplanet KIC~1255b,
which has the best quality data of the three candidates.
In principle, the model can be applied to the other two candidate evaporating rocky exoplanets
after some additional work.
Specifically, for \mbox{KOI-2700b} it would be necessary to obtain a better constraint on the dust survival time
from possible correlations between subsequent transits or lack thereof.
Modelling \mbox{K2-22b} would require the initial launch velocity of the grains
to be explored in more detail (see footnote~\ref{fn:k2_22b}).

The basic parameters of the KIC~1255b system are listed in Table~\ref{tbl:sys_pars}.
These are the values that we use in calculations throughout the rest of this paper.
For the stellar parameters, there are different estimates in the literature,
casting doubt on whether the star has evolved off the main sequence or not.
In Appendix~\ref{app:star_param},
we investigate the different claims
and conclude that the star is most likely still on the main sequence.

The rest of this paper is organised as follows.
Section~\ref{s:meth} provides a detailed description of the dust cloud model and of how it is compared to the observations.
Section~\ref{s:res} gives the resulting constraints on the free parameters of the model
and shows what they imply for the dust composition.
In Sect.~\ref{s:disc}, we discuss our findings in the light of previous work
and examine one of our modelling assumptions.
Finally, we summarise our work and draw conclusions in Sect.~\ref{s:conclusions}.


\section{Methods}
\label{s:meth}

In our modelling efforts, we adopt the following approach.
First, we calculate the shape of the dust tail
by following dust grains from their release from the planet to their complete sublimation (Sect.~\ref{s:meth_eom}).
We then generate the light curve that would result from the transit of this dust tail (Sect.~\ref{s:meth_lc}).
Finally, the synthetic light curve is compared to the phase-folded \textit{Kepler} data, yielding a goodness of fit (Sect.~\ref{s:meth_gof}).
These three steps are carried out for different values of input parameters
(which include the material properties of the dust)
in a Markov chain Monte Carlo (MCMC) framework
to obtain constraints on those parameters (Sect.~\ref{s:meth_frame}).

\begin{table}[!t]
  \centering
  \small
  \caption{Modelling assumptions}
  \label{tbl:assump}
  {
  \renewcommand{\arraystretch}{1.15}
  \begin{tabular}{clcc}
  \hline
  \hline
  Nr. & Assumption & Used & Notes \\
  \hline
   1 & Steady state & \checkmark & \\
   2 & Optically thin dust cloud & \checkmark & (a) \\
   3 & Single initial grain size & \checkmark & \\
   4 & Spherical dust grains & \checkmark & \\
   5 & Constant optical efficiency factors & & \\
   6 & Long dust survival time & & (b) \\
   7 & Constant dust sublimation rate & & \\
   8 & Dust sublimation as in vacuum & \checkmark & \\
   9 & Circular planet orbit & \checkmark & \\
  10 & Small planet & \checkmark & \\
  11 & Radiation controlled dust temperatures  & \checkmark & \\
  12 & Radiation pressure dominated dynamics & \checkmark & \\
  \hline
  \end{tabular}
  \tablefoot{
  An overview of the modelling assumptions introduced in \citetalias{2014A&A...572A..76V}.
  The check marks indicate whether these assumptions are still used in the present work. \\
  \tablefoottext{a}{This assumption is investigated in more detail in Sect.~\ref{s:disc_optical_depth}.} \\
  \tablefoottext{b}{In \citetalias{2014A&A...572A..76V},
  the dust survival time was assumed to be long (compared to the orbital period of the dust grains).
  Here, we assume instead that it is short
  (not much longer than one orbital period of the planet;
  see Sect.~\ref{s:meth_t_surv} for further details).}
  }
  }
\end{table}

Our model makes use of a number of assumptions,
which were introduced in \citetalias{2014A&A...572A..76V} (see Table~\ref{tbl:assump}).
They are motivated in the rest of this section when they are encountered
and in Sect.~4.1 of \citetalias{2014A&A...572A..76V}.
The numerical approach of this work allows us to
let go of several of the assumptions made in \citetalias{2014A&A...572A..76V}:
The optical efficiency factors of the dust (e.g., $ Q\sub{abs} $, $ Q\sub{pr} $) are allowed to change as dust grains become smaller (assumption~5).
We do not use orbit-averaged quantities,
and hence the survival time of the dust grains is not required to be long compared to the orbital period of the dust grains (assumption~6).
Sublimation rates of dust grains are calculated in a time-dependent manner, rather than assumed to remain equal to their initial value (assumption~7).

In contrast to \citetalias{2014A&A...572A..76V}, we do not primarily test several possible dust species
for which detailed laboratory measurements of their properties are available,
but rather describe the optical and thermodynamic properties of the dust material with a set of free parameters.
The constraints on these parameters obtained from the MCMC fitting
can then be compared with laboratory measurements of candidate dust species.

\subsection{Tail morphology}
\label{s:meth_eom}

To determine the shape of the dust tail,
we compute the trajectory of a single dust grain.
Assuming that all dust grains released from the planet are identical
(i.e., they have the same composition and initial size),
this trajectory gives the shape of a stream of dust particles,
with individual time steps corresponding to separate particles launched from the planet at different times.
Although the single-initial-grain-size assumption makes our model somewhat less realistic,
it lowers the computation time of the model significantly,
allowing us to calculate many different model realisations in a statistical fashion.
The initial size of the dust grains that we eventually find
should be interpreted as a typical initial size.
In addition, we note that the origin of the dust grains through condensation in a planetary outflow may favour a narrow size distribution,
since large grains cannot be lifted out of the atmosphere \citep{2013MNRAS.433.2294P}
and the saturated-atmosphere crossing time could result in a minimum size.

The path of an individual dust grain is determined by solving the equations of motion and sublimation.
These equations are coupled because
(1)~the grain-size evolution influences the dust dynamics through size-dependent radiation pressure,
and (2)~the dynamics influence the sublimation rate through distance-dependent grain temperatures.

\subsubsection{Dust dynamics}
\label{s:meth_eom_dyn}

After being released, the dust particles drift away from the planet
as a result of the direct radiation pressure force of the star.
Aside from the stellar gravity, we assume this to be the only relevant force for the dynamics of the dust grains.
We ignore the gravitational influence of the planet,
which is only important within the planet's Roche lobe,
i.e., at radii of at most about 1~R$ \sub{\oplus} $ from the planet's centre
(calculated from the upper limit on the planet's mass of 0.02~M$\sub{\oplus}$ of \citealt{2013MNRAS.433.2294P}).
We also neglect
Poynting--Robertson drag (only relevant over many orbits),
stellar wind pressure \citep[see Appendix~A of][]{2014ApJ...784...40R},
and gas drag from the planetary outflow (assumed to diminish rapidly).
Because the radiation-pressure-induced drift is slow with respect to the local Keplerian velocity,
the dynamics are best solved in a rotating reference frame
(i.e., centred on the star and co-rotating with the planet).
Hence, the motion of a dust grain is described by
\begin{equation}
  \label{eq:eom1}
  \frac{ \dif^2 \vec{r} }{ \dif t^2 }
    = - \underbrace{ \frac{ G M\sub{\star} ( 1 - \beta ) }{ r^3 } \vec{r} }_{ \substack{ \text{Gravity and} \\ \text{radiation pressure} } }
      - \underbrace{ 2 \boldsymbol{ \upomega } \times \frac{ \dif \vec{r} }{ \dif t } }_{ \substack{ \text{\vphantom{g}Coriolis} \\ \text{acceleration} } }
      - \underbrace{ \vphantom{\frac{a}{b}} \boldsymbol{ \upomega } \times ( \boldsymbol{ \upomega } \times \vec{r} ) }_{ \substack{ \text{Centrifugal} \\ \text{acceleration} } }.
\end{equation}
Here,
the vector $ \vec{r} $ is the position of the particle
(and hence its magnitude $ r $ is the distance to the centre of the star),
$ t $ denotes time,
$ G $ is the gravitational constant,
$ \beta $ is the ratio between the norms of the direct radiation pressure force and the gravitational force,
and $ \boldsymbol{ \upomega } $ is the rotation vector of the reference frame
(with magnitude $ \omega\sub{p} = 2 \pi / P\sub{p} $).
The Coriolis and centrifugal acceleration terms
represent fictitious forces due to the rotating reference frame.

For spherical dust grains, the $ \beta $ ratio is given by \citep[e.g.,][]{1979Icar...40....1B}
\begin{equation}
  \label{eq:beta}
  \beta
    = \frac{ 3 }{ 16 \pi c G }
      \frac{ L\sub{\star} }{ M\sub{\star} }
      \frac{ \bar{Q}\sub{pr}( s ) }{ \rho\sub{d} s }.
\end{equation}
Here,
$ c $ is the speed of light,
$ \bar{Q}\sub{pr} $ is the radiation pressure efficiency averaged over the stellar spectrum (see Sect.~\ref{s:meth_dust_prop}),
$ \rho\sub{d} $ is the density of the dust material,
and $ s $ is the grain radius.

\subsubsection{Dust sublimation}

For a spherical dust grain in a gas-free environment,
the rate at which the grain radius $ s $ changes is given by
(\citealt{1913PhRv....2..329L}; see also Eq.~(11) of \citetalias{2014A&A...572A..76V})
\begin{equation}
  \label{eq:dsdt}
  \frac{ \dif s }{ \dif t }
    = - \frac{ J(T\sub{d}) }{ \rho\sub{d} }
    = - \frac{ \alpha p\sub{v}(T\sub{d}) }{ \rho\sub{d} }
    \sqrt{ \frac{ \mu m\sub{u} }{ 2 \pi k\sub{B} T\sub{d} } }.
\end{equation}
Here,
$ J $ is the mass loss rate per unit surface (units: [g~cm$^{-2}$~s$^{-1}$]; positive for mass loss),
$ \alpha $ is the evaporation coefficient
(which parametrises kinetic inhibition of the sublimation process),
$ p\sub{v} $ is the partial vapour pressure at phase equilibrium,
$ \mu $ is the molecular weight of the molecules comprising the dust, $ m\sub{u} $ is the atomic mass unit,
$ k\sub{B} $ is the Boltzmann constant, and $ T\sub{d} $ is the temperature of the dust.
The equilibrium vapour pressure
is material-specific
and depends strongly on temperature $ T $.
This material and temperature dependence is captured by
the parameters $ \mathcal{A} $ and $ \mathcal{B} $
in the Clausius--Clapeyron relation
\begin{equation}
  \label{eq:pres_vap}
  p\sub{v}( T ) = \exp ( \, - \mathcal{A} / T + \mathcal{B} \, ) \; \mathrm{ dyn~cm^{-2} }.
\end{equation}
We assume $ \alpha $, $ \mathcal{A} $, and $ \mathcal{B} $ to be independent of temperature
and ignore any temperature dependence of $ p\sub{v} $ beyond that given by Eq.~\eqref{eq:pres_vap}.

Without changing the temperature dependence of the sublimation rate
(i.e., without changing the functional form of equation \eqref{eq:dsdt}),
the parameters $ \alpha $, $ \mu $, and $ \mathcal{B} $
can be combined into the new parameter\footnote{%
Generally, the difference between $ \mathcal{B} $ and $ \mathcal{B}' $
is relatively small (because $ \alpha $ and $ \mu $ affect $ \mathcal{B}' $ with opposite sign).
For the dust species listed in Table~3 of \citetalias{2014A&A...572A..76V},
$ 0.01 \lesssim \alpha \lesssim 1 $, $ 10 \lesssim \mu \lesssim 200 $,
and $ | \, \mathcal{B} - \mathcal{B}' \, | / \mathcal{B} \lesssim 7 \% $,
which is comparable to the typical uncertainties on $ \mathcal{B} $.}
\begin{equation}
  \label{eq:b_prime}
  \mathcal{B}' \equiv \mathcal{B} + \ln( \alpha \sqrt{\mu} ).
\end{equation}
This reduces the number of material-specific free parameters of our model,
with the sublimation properties of dust materials
now fully described by the free parameters $ \mathcal{A} $ and $ \mathcal{B}' $.

Dust grain temperatures $ T\sub{d}( s, r ) $ are calculated from the power balance
between incoming stellar radiation and outgoing thermal radiation.
This ignores the latent heat of sublimation and the collisional heating by stellar wind particles,
which are both insignificant \citep{1974A&A....35..197L,2014ApJ...784...40R}.
The power balance reads
\begin{equation}
  \label{eq:temp_d}
  \frac{ \Omega( r ) }{ \pi } \! \! \int \! Q\sub{abs}( s, \lambda ) \, F_{\lambda,\star}( \lambda ) \, \dif \lambda
    = 4 \pi \! \int \! Q\sub{abs}( s, \lambda ) \, B_\lambda( \lambda, T\sub{d} ) \, \dif \lambda,
\end{equation}
where
$ \lambda $ denotes wavelength,
$ Q\sub{abs} $ is the monochromatic absorption efficiency of the dust grain,
$ F_{\lambda,\star} $ is the star's flux at the stellar surface
(units: [erg~s$^{-1}$~cm$^{-2}$~cm$^{-1}$]),
$ B_\lambda $ denotes the Planck function,
and $ \Omega( r ) $ is the solid angle subtended by the star as seen from the dust particle.
The distance dependence is given by
$ \Omega( r ) = 2 \pi \left[ 1 - \sqrt{ 1 - ( R\sub{\star} / r )^2 } \, \right] $.
For the stellar spectrum of KIC~12557548 we take a \citet{1993KurCD..13.....K} model
with $ T\sub{eff,\star} = 4500~\mathrm{K} $ and a surface gravity of $ \log g  = 4.5 $ (see Fig.~\ref{fig:lnk}),
scaled with the stellar luminosity.
Equation~\eqref{eq:temp_d} is solved numerically for $ T\sub{d} $ as a function of grain size and distance from the star.

\subsubsection{Optical properties of the dust}
\label{s:meth_dust_prop}

Several parts of our model
(the calculation of $ \beta $ and $ T\sub{d} $, as well as the synthetic-light-curve generation)
require values for the dimensionless efficiency factors
$ Q\sub{abs} $, $ Q\sub{ext} $, $ Q\sub{sca} $, and $ Q\sub{pr} $
with which dust particles interact with the stellar radiation field
as a function of grain size and wavelength.
We assume the dust grains to be solid spheres,
which allows us to calculate these quantities using \citet{1908AnP...330..377M} theory.
As input, the Mie calculations require (material dependent) complex refractive indices as a function of wavelength.
The complex refractive index consists of a real part $ n ( \lambda ) $ and an imaginary part $ k ( \lambda ) $.

To allow our model to retrieve the dust composition in a relatively unbiased manner,
we wish to describe the complex refractive index in a parametric way,
rather than testing a limited number of dust species with full wavelength-dependent $ n ( \lambda ) $ and  $ k ( \lambda ) $ data.
The parametrisation should be simple enough to be numerically feasible,
but still able to capture the optical properties of a wide range of materials.
In particular, the dependence of dust temperature on grain size should be described well,
since the sublimation of dust grains is very sensitive to it.
We tested several different prescriptions for the wavelength dependence of
the complex refractive index and finally settled on a very simple recipe,
in which $ n ( \lambda ) $ is assumed to be constant over all wavelengths
and $ k ( \lambda ) $ is split into two constants for different wavelength regimes:
\begin{equation}
  \label{eq:lnk_recipe}
  n ( \lambda ) = n, \qquad k ( \lambda ) =
  \begin{cases}
     k_1 & \text{for } \lambda < \lambda\sub{split} \\
     k_2 & \text{for } \lambda \geq \lambda\sub{split}
  \end{cases}\!.
\end{equation}
In our model, $ n $ and $ k_1 $ are free parameters,
while $ k_2 $ and $ \lambda\sub{split} $ are fixed at $ k_2 = 1 $ and $ \lambda\sub{split} = 8 \; \upmu$m.
The values we chose for $ \lambda\sub{split} $ and $ k_2 $ reflect
the fact that many dust species have strong features in their $ k ( \lambda ) $ profile
beyond $ 8 \; \upmu $m,
which can facilitate their cooling
if their heating efficiency is sufficiently low
(see Appendix~\ref{app:lnk_recipe}).

The main advantage of our simple complex-refractive-index recipe is the limited number of free parameters it uses.\footnote{%
More complex prescriptions for the optical properties, with more free parameters
(e.g., with $ k_2 $ and/or $ \lambda\sub{split} $ kept free, or with $ k ( \lambda ) $ split up into more wavelength regimes),
pose computational difficulties to the MCMC analysis,
which for increasing dimensionality needs more iterations to reach convergence and obtain reliable parameter constrains.
Conversely, an even more simple recipe, using a constant $ k $ for all wavelengths,
fails to capture the cooling through mid-infrared features,
giving relatively high dust temperatures even for $ k \sim 10^{-5} $.}
Despite its simplicity, however, the recipe
is capable of reproducing much of the diversity seen in grain temperatures
seen for real dust species.
This is demonstrated in Appendix~\ref{app:lnk_recipe},
where we compare the grain-size-dependent dust temperatures predicted by our simple recipe to those found
using the full wavelength-dependent complex-refractive-index data of real dust species.

\subsubsection{Numerical method}

To determine the path and size-evolution of a dust grain,
we integrate Eqs.~\eqref{eq:eom1} and \eqref{eq:dsdt} using a classical fourth order Runge--Kutta scheme.
The dust properties that couple these equations, $ \beta ( s ) $ and $ T\sub{d} ( s, r ) $,
are precalculated and tabulated.
The initial size of the dust grain $ s\sub{0} $ is a free parameter of our model;
the other initial conditions (position and velocity) are set to simulate a particle
released at the position of the planet with zero velocity with respect to the planet.
This means that we neglect the velocity with which particles are launched away from the planet by the gas outflow,
assuming that radiation pressure dominates the dynamics of the dust grains.
For KIC~1255b, this seems to be a good approximation
(see Sect.~4.1 of \citetalias{2014A&A...572A..76V} and Sect.~6.2 of \citealt{2015ApJ...812..112S}).
By releasing the particles from a single point (corresponding to the centre of the planet),
we also ignore the fact that real dust grains will have a range of starting positions,
due to the non-zero size of the planet, interactions with the planet's gas outflow, and the planet's gravity.
These processes all happen within the Roche lobe of the planet ($ r \lesssim 1$~R$\sub{\oplus} $),
which our model treats as a ``black box'', focussing on the larger-scale structure dust tail.
During the integration, a dynamical step-size control ensures that
(1)~the position of the particle (in the rotating frame) never changes by more than
1\% of the stellar radius
(mostly to ensure an adequate spatial sampling from which a light curve can be generated)
and (2)~its size does not decrease by more than a factor 2.
The integrator stops when a particle has reached a size of $ s \leq 0.001 $~$\upmu$m.
At this point the particle is considered fully sublimated
as it does not contribute significantly anymore to the extinction or scattering of star light.

\begin{figure}[!t]
  \includegraphics[width=\linewidth]{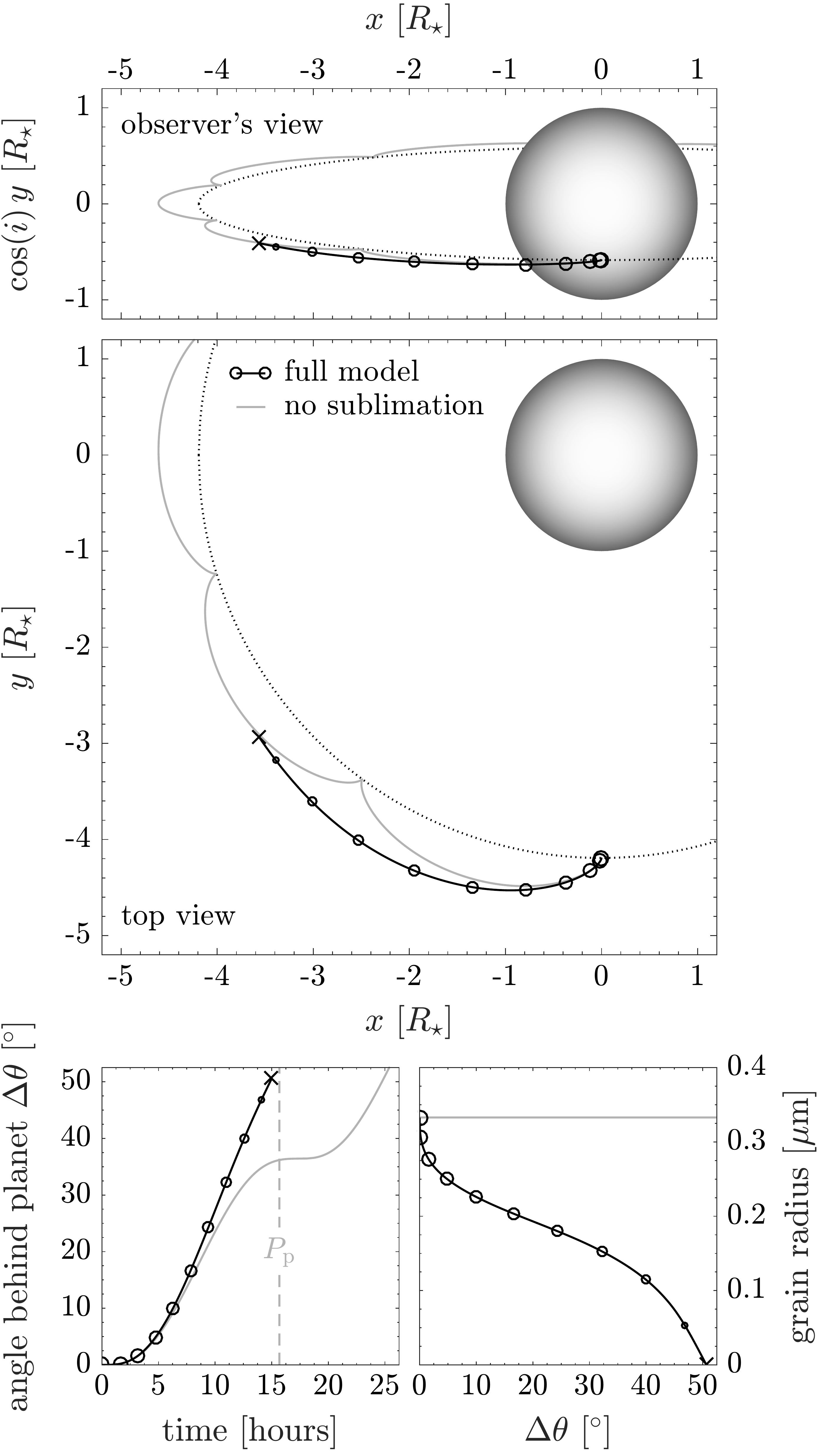}
  \caption{
  Example dust tail morphology of KIC~1255b
  found by our numerical dynamics and sublimation calculation (black lines).
  The \mbox{$\ocircle$-symbols} shown at several points along the stream
  are placed at equal intervals of travel time.
  Their size is proportional to the local grain size.
  An \mbox{$\times$-symbol} marks the point where the dust grains are fully sublimated.
  For comparison, a dust path ignoring sublimation is shown in grey.
  Several orbits of the dust grain are shown,
  which results in a rosette-like shape in the corotating frame (see, also, Fig.~1 of \citetalias{2014A&A...572A..76V}).
  The dotted line indicates the orbit of the planet.
  \textbf{Top:}~The dust tail and host star
  from the observer's point of view
  at the time of mid-transit.
  \textbf{Middle:}~View looking down onto the planet's orbital plane.
  \textbf{Bottom left:}~Angle between a dust grain and the planet as a function of time.
  The vertical dashed line indicates the orbital period of the planet~$ P\sub{p} $.
  \textbf{Bottom right:}~Grain size as a function of angle behind the planet $ \Delta\theta $.
  }
  \label{fig:eom_example}
\end{figure}

The output of this dynamics-and-sublimation routine is a list of particle positions and sizes at each time step.
Assuming that the planet ejects a continuous stream of particles with identical composition and initial size,
this list corresponds to the coordinates of particles that have left the planet at different points in time.
The model solutions presented in this work
typically consist of several hundred output points.
Figure~\ref{fig:eom_example} shows an illustrative example of a dust stream predicted by our model,
including the evolution of the grain size.
It is compared to a dust path that does not take into account the size evolution due to sublimation.

The dust dynamics are solved in the orbital plane of the planet.
In order to find the coordinates with respect to the star as seen from the observer (top panel in Fig.~\ref{fig:eom_example}),
the output coordinates are rotated by an angle $ ( 90\mdegr - i ) $.
Here, $ i $ denotes the planet's orbital inclination,
which we compute from the transit impact parameter $ b $ (a free parameter of our model)
using $ \cos i = b R\sub{\star} / a\sub{p} $.

\subsection{Light-curve generation}
\label{s:meth_lc}

The previous step gives the spatial and size distribution of dust
in the form of grain sizes and three-dimensional position coordinates at each time step,
which correspond to coordinates and sizes of dust particles released from the planet at different times.
From these data,
we now have to compute the normalised flux into the direction of the observer as a function of phase.
Rather than determining spatial densities of dust grains,
we compute the effect on the light curve of each individual particle in the dust stream.
After computing the effects on the light curve of a single dust grain,
these are scaled using the dust mass loss rate of the planet $ \dot{M}\sub{d} $,
a free parameter of the model.

At each step in phase,
we determine which particles are in front of the star and which are behind the star
as seen from the observer.
Any particle that is in front of the star reduces the flux because of extinction.
This is achieved by subtracting
an amount of flux proportional to the grain's extinction cross-section and the local intensity of the stellar disk.
To describe the stellar intensity profile,
we employ the four-term limb-darkening law of \citet{2011A&A...529A..75C}
with parameters
$ a_1 = 0.71 $, $ a_2 = -0.83 $, $ a_3 = 1.52 $, $ a_4 = -0.56 $,
appropriate for a star with $ T\sub{eff,\star} = 4500~\mathrm{K} $ and $ \log g  = 4.5 $,
observed in the \textit{Kepler} band.

Any particles that are not behind the star increase the flux through scattering.
The amount of flux to add is proportional to the grain's scattering cross-section,
their scattering phase function at the scattering angle, and the intensity of the stellar disk at the point from where the light emanates.
For the scattering, it is important to take into account the non-zero size of the star
\citep[see, also,][]{2015MNRAS.454....2B,2016MNRAS.461.2453D}.
This is especially true for grains with sharply forward peaked phase functions.
Rather than convolving the phase function with the angular size of the star as seen from the dust cloud,
we integrate over the limb-darkened stellar disk using Monte Carlo integration
(taking into account the solid angle subtended by the star as seen from the dust particle).

\begin{figure}[!t]
  \includegraphics[width=\linewidth]{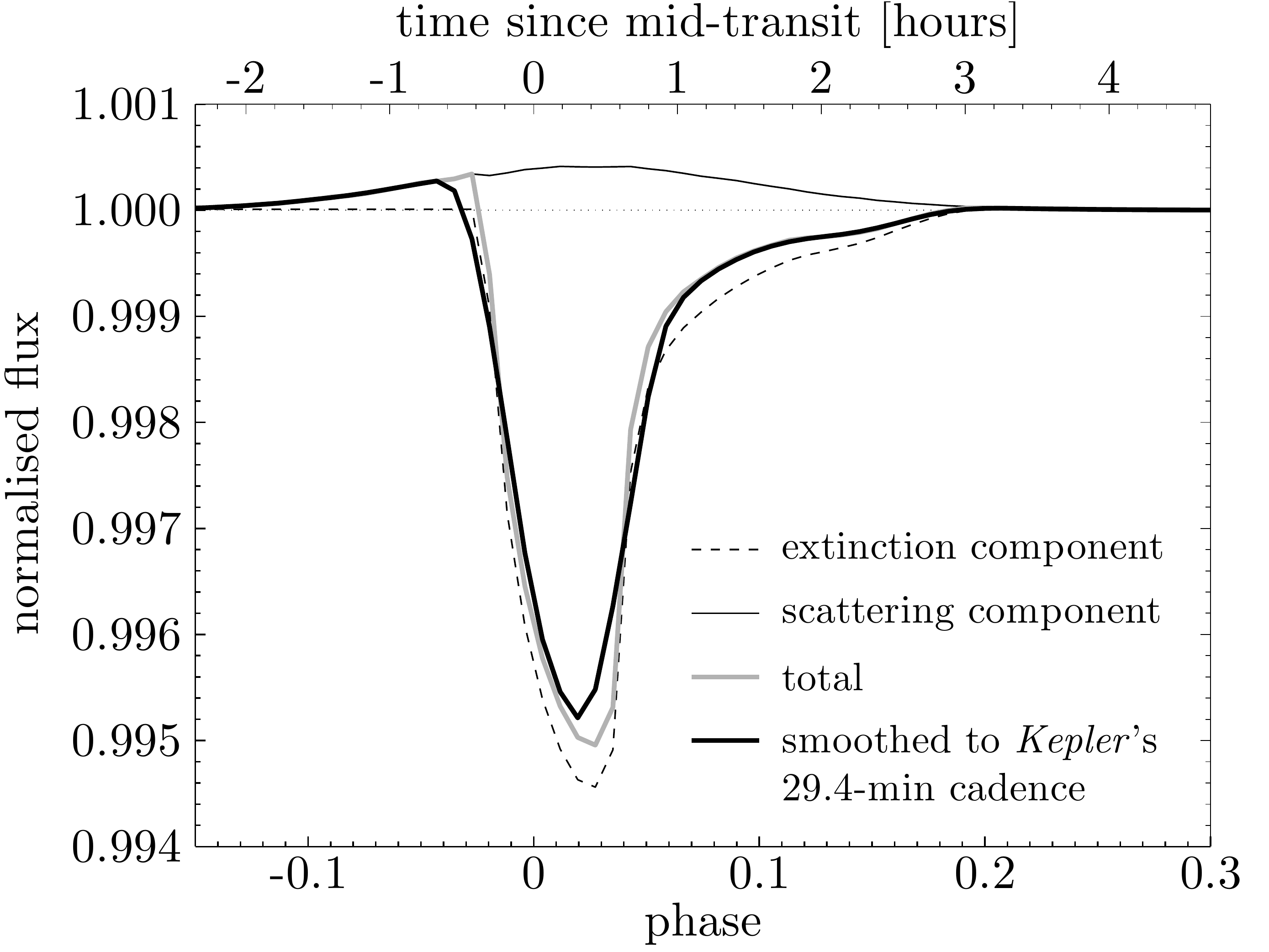}
  \caption{
  Example transit light curve of KIC~1255b generated by our model,
  together with the isolated extinction and scattering components.
  Note that while this figure only shows the part of the light curve around the transit,
  we calculate fluxes at all orbital phases.
  }
  \label{fig:lc_example}
\end{figure}

The extinction and scattering components of the light curve are combined
into a synthetic light curve (see Fig.~\ref{fig:lc_example}).
Three final operations need to be done on the synthetic light curve
before it can be compared with the observations.
First, the light curve is convolved with a trapezoidal kernel
to capture the combined effects of
the 29.4~minute exposure time of the long cadence \textit{Kepler} data
and the binning of a phase-folded light curve.\footnote{%
The kernel has a midsegment length (i.e., full width at half maximum)
corresponding to the long-cadence exposure time,
and ramp-up and ramp-down lengths corresponding to the bin width.}
Since the exposure time is longer than the timescale on which the light curve varies,
this introduces significant smoothing in the light curve.
Second, we normalise the light curve to the out-of-transit flux
in the same way as the \textit{Kepler} data were normalised.
In principle, this may affect model realisations that show significant scattering outside the in-transit phase window,
but in practice we find that this procedure makes little difference.
Third, the synthetic light curve is shifted in phase by an amount $ \Delta\varphi\sub{0} $,
a free parameter of the model that reflects the unknown mid-transit time of the planet.

We do not take into account the effect on the light curve of the solid planet.
Given the upper limits on the size of the planet,
\citep[$ R\sub{p} \lesssim 1 $~R$\sub{\oplus} $;][]{2014A&A...561A...3V}
it causes a flux deficit of the order of $ ( R\sub{p} / R\sub{\star} )^2 \lesssim 10^{-4} $,
whereas the average transit depth measured is about 0.5\%.
In principle, the planet's size (or that of any optically thick part of the dust cloud)
could be added to the model as an additional free parameter.
However, since the effect of the planet is small, and to avoid further complexity of the model,
we choose to ignore it.

\begin{table*}[!t]
  \centering
  \small
  \caption{Free parameters of the model}
  \label{tbl:free_pars}
  {
  \renewcommand{\arraystretch}{1.15}
  \begin{tabular}{lcccc}
  \hline
  \hline
  Parameter & Symbol & Units & Scale & Bounds \\
  \hline
  Transit midpoint phase offset & $ \Delta\varphi\sub{0} $ & & linear & $ [ -0.5, 0.5 ] $ \\
  Transit impact parameter & $ b $ & & linear & $ [ 0, 1 ] $ \\
  Dust mass loss rate of planet & $ \dot{M}\sub{d} $ & M$\sub{\oplus}$~Gyr$^{-1}$ & logarithmic & $ [ 0, +\infty ) $\\
  Initial grain size & $ s\sub{0} $ & $\upmu$m & logarithmic & $ [ 0.01, +\infty ) $ \\
  Material density of the dust grains & $ \rho\sub{d} $ & g~cm$^{-3}$ & logarithmic & $ [ 0.01, 10 ] $ \\
  Real part of complex refractive index & $ n $ & & linear & $ [ 1, 4 ] $ \\
  Imaginary part of complex refractive index for $ \lambda < 8 $~$\upmu$m & $ k_1 $ & & logarithmic & $ [ 10^{-5}, 10 ] $ \\
  Sublimation parameter 1 & $ \mathcal{A} $ & $ 10^{4} $ K & linear & $ [ 4, 10 ] $ \\
  Sublimation parameter 2 & $ \mathcal{B}' $ & & linear & $ [ 20, 50 ] $ \\
  \hline
  \end{tabular}
  }
\end{table*}

\subsection{Goodness-of-fit evaluation}
\label{s:meth_gof}

We now describe the observational data and
how the goodness of fit of model light curves is calculated.
In principle, determining the relative likelihood of a particular set of parameter values
is done by a simple $ \chi^2 $ comparison of the synthetic light curve with the observed one.
However, we use several additional criteria to assess the viability of the model solution
before the $ \chi^2 $ statistic is computed.
If a given model solution does not satisfy any of these criteria,
the likelihood of this solution is set to zero.

\subsubsection{Observations}
\label{s:meth_obs}

We compare the model solutions with the \textit{Kepler} light curve of KIC~1255b.
Specifically, we use the long-cadence data of quarters 1 through 15,
as reduced by \citet[see their Sect.~2]{2014A&A...561A...3V},
which we phase folded and binned into 128 phase bins.
The length of one bin corresponds to about 0.25 of the \textit{Kepler} long-cadence integration time.
Figure~\ref{fig:intro_sketch} shows the resulting light curve.

Uncertainties on the light curve were determined in two different ways.
First, we compute the standard deviation of the individual \textit{Kepler} data points within each phase bin.
This results in phase-dependent error bars.
Because of the orbit-to-orbit variability in the transit depth,
the in-transit part of the light curve has greater uncertainties than the out-of-transit part.
The variability-caused variance is roughly proportional to the local (i.e., local in phase) transit depth.
In a second approach, we estimated the average error on the binned data
from the variance in normalised flux amongst the out-of-transit ($ \varphi \in [-0.5, -0.15] \cup [0.2, 0.5] $) bins.\footnote{%
This assumes that the out-of-transit part of the light curve is flat,
which was demonstrated by \citet[see their Sect.~3.3]{2014A&A...561A...3V}.
It gives a good estimate of the actual uncertainties on the data,
less affected by noise that is correlated on timescales longer than the length of an individual bin (i.e., red noise).}
This turned out to be a factor of about 1.4 greater than the median of the uncertainties found from the standard deviation within each bin.
The latter were therefore multiplied by this factor to get the final uncertainties shown in Fig.~\ref{fig:intro_sketch} and used for our fitting.

\subsubsection{Dust-survival-time constraints}
\label{s:meth_t_surv}

The full \textit{Kepler} light curve contains information about the system
that is not preserved in the phase-folding process.
Specifically, the variability of the transit depths is sensitive to the survival time of the dust grains $ t\sub{surv} $.
Correlations (or the lack thereof) between the depths of subsequent transits
or the depth of a transit core and the egress depth of the following transit
can be used to determine how long after their release from the planet dust grains influence the transit depth.
In their thorough analysis of the long-cadence \textit{Kepler} data,
\citet{2014A&A...561A...3V} found no evidence of
any such correlations.
Their absence
indicates that the survival time of the dust grains is not much longer than the planet's orbital period $ P\sub{p} $.

To use this information, we set a maximum survival time for the dust grains, $ t\sub{surv} < t\sub{cut} $.
This is numerically convenient,
because it severely restricts the maximum computation time for a given model evaluation.
The maximum survival time is set to $ t\sub{cut} = 1.2 \, P\sub{p} $ rather than $ 1.0 \, P\sub{p} $ (orbital period of the planet)
to give some leeway to solutions in which dust grains survive for slightly longer than $ 1.0 \, P\sub{p} $,
but with the last part contributing very little to the extinction,
such that no significant correlations are produced between subsequent transits.\footnote{%
A more thorough way of using the absence of correlation between subsequent transits would be
to simulate a varying mass loss rate and inspect the resulting correlation between transits,
or to compute the light curve for material older than $ 1.0 \, P\sub{p} $ and compare this to switch-off events in the observed light curve.
However, these methods come with prohibitively large computational costs
and are therefore beyond the scope of the present work.}
We find that the posterior probability distribution of survival time peaks around $ 1.0 \, P\sub{p} $,
confirming that the cut-off time of $ 1.2 \, P\sub{p} $ is reasonable.

\subsubsection{Light-curve criteria}
\label{s:meth_lc_crit}

Our model is very sensitive to some of the input parameters.
In particular, the dust sublimation rate depends strongly on temperature
and, as a result, small changes in the values of the material properties can result in large changes in the light curve.
To prevent the fitting algorithm from wasting computational resources on
parts of the parameter space that produce light curves that are clearly unrealistic,
we only allow solutions that
display the general features of the light curve,
i.e., the asymmetric shape with a gradual egress and a pre-ingress brightening.
Technically, this is achieved by the following three cumulative criteria:
\begin{enumerate}
  \item Only model solutions that are within $20 \sigma$ of the observational data at all phases are allowed.
  This excludes synthetic light curves without a main transit feature,
  or with (additional) strong transit features at other phases than the observed one.
  \item At phases $ \varphi \in [0.08, 0.10] $, we allow a maximum normalised flux of $ 0.9999 $.
  This excludes models without a significant tail.
  \item At phases $ \varphi \in [-0.05, -0.04] $, we require a minimum normalised flux of $ 1.0001 $.
  This excludes models that do not exhibit any pre-ingress brightening.
\end{enumerate}
We acknowledge that this procedure may in principle lead to an underestimation of the uncertainties we derive on the free parameters,
because it will reject some model solutions that are very unlikely, but still have a non-zero likelihood.

\subsection{Fitting strategy}
\label{s:meth_frame}

Our dust tail model consists of three steps:
(1)~determining the shape of the dust cloud,
(2)~computing a synthetic light curve,
and (3)~comparing it to the observations.
In order to put constraints on the dust composition,
these steps need to be repeated many times for different values of the free parameters.
We now describe the parameter space
that needs to be explored
and the method we use to do so.

\subsubsection{Free parameters}
\label{s:meth_pars}

Our model contains nine free parameters,
five of which describe the dust material ($ \rho\sub{d} $, $ n $, $ k_1 $, $ \mathcal{A} $, and  $ \mathcal{B}' $).
They are listed in Table~\ref{tbl:free_pars},
along with their scaling and the bounds of the range considered for each parameter.
We assume flat or log flat prior probability distributions for all free parameters within their allowed ranges.
In principle, the stellar parameters could be added as free parameters
to account for their uncertainty,
but given the already large number of free parameters, we keep them fixed
at the fiducial values listed in Table~\ref{tbl:sys_pars}.

For $ \Delta\varphi\sub{0} $, $ b $, and $ \dot{M}\sub{d} $, we allow the entire physically possible range.
For the other parameters, the bounds reflect physically reasonable limits and sometimes numerical limitations.
The initial grain size $ s\sub{0} $ and material density $ \rho\sub{d} $
are required to be higher than 0.01~$\upmu$m and 0.01~g~cm$^{-3}$, respectively,
primarily for numerical reasons.
Very-low-density grains have high \mbox{$\beta$ ratios} (see Eq.~\eqref{eq:beta})
and are put on unbound trajectories that require long computation times, but do not produce good fits.
The upper bound on $ \rho\sub{d} $ is set to 10~g~cm$^{-3}$.
This is high enough to consider dust grains made of pure iron,
while only relatively rare metals such as lead and gold have even higher densities.
For the complex-refractive-index and sublimation parameters ($ n $, $ k_1 $, $ \mathcal{A} $, and  $ \mathcal{B}' $),
the bounds are set to bracket the values found by laboratory measurements for a wide range of possible dust species.

\subsubsection{Markov chain Monte Carlo method}
\label{s:meth_mcmc}
 
The parameter space we have to consider is very large.
On average, a single model evaluation takes about 1\,s of computation time on a desktop workstation,
making it unfeasible to search for maxima in the likelihood using a grid approach.
Therefore, we employ an MCMC method to explore the parameter space.
Specifically, we use the affine-invariant ensemble-sampler algorithm of \citet{goodman2010}
as implemented in the Python package \texttt{emcee} \citep{2013PASP..125..306F}.
This algorithm is designed to efficiently sample probability distributions with strong correlations between parameters.
It uses a ensemble of ``walkers'' that map out the probability density landscape
by moving through parameter space.
Their proposed steps are based on the positions of the other walkers in the ensemble
and the acceptance of a step depends on the probability ratio of the proposed and current positions in parameter space.

We use 100 walkers, initialised
at positions in parameter space that were found to give good fits in earlier trial runs.
The proposal-scaling factor \citep[the parameter $ a $ in][]{goodman2010} is set at its recommended standard value of 2.
We find that the acceptance rate of proposed steps is rather low (about 6 to 7\%),
indicating bad mixing of the chain.
Increasing the number of walkers only marginally improves the acceptance rate,
while the associated computational costs rapidly become prohibitively high.
Hence, we compensated the low acceptance rate with a large number of steps.
After burn-in,
we let the walkers make $ 9 \times 10^4 $ steps each,
resulting in a total of $ 9 \times 10^6 $ model evaluations.
We assessed the convergence of the chain using the autocorrelation length,
which is found to be less than 5000 steps for all free parameters.
This means that the length of the chain is longer than 18 autocorrelation times for all parameters
(\citealt{2013PASP..125..306F} recommend a chain length of 10 autocorrelation times)
and we can assume the chain to be converged.
The number of unique model solutions in the flattened chain is about $ 6 \times 10^5 $.
We find that the nine-dimensional posterior probability density function can be mapped out in sufficient detail with this number of samples.


\section{Results}
\label{s:res}

\begin{figure}
  \includegraphics[width=\linewidth]{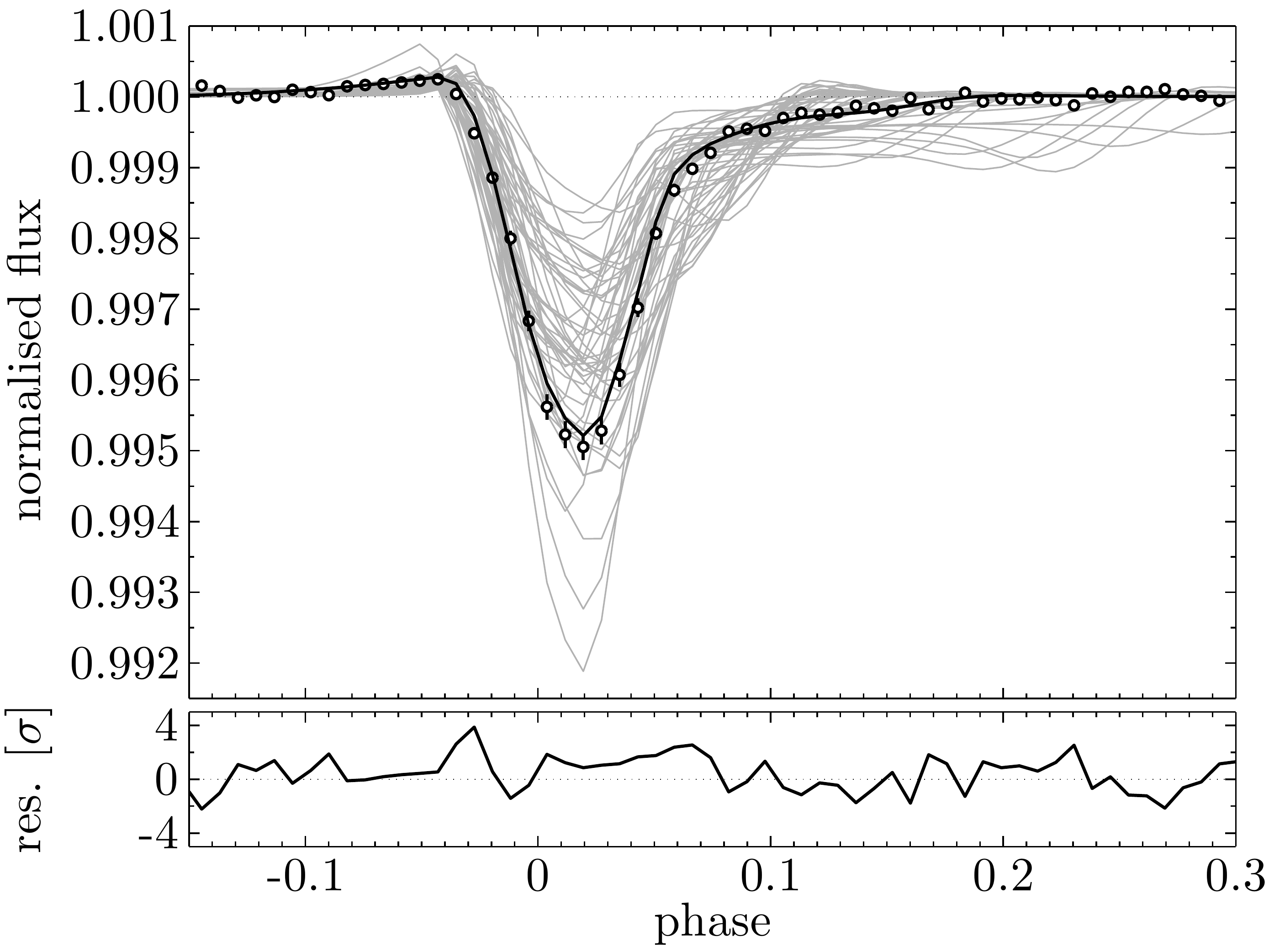}
  \caption{
  \textbf{Top:}~Comparison between the best-fit model (black line) and the \textit{Kepler} data (black circles with error bars).
  To show the range of variation within the Markov chain, 50 randomly selected samples from the chain are underplotted as grey lines.
  \textbf{Bottom:}~Residuals of the best-fit model, normalised using the phase-dependent error bars on the data.
  }
  \label{fig:mdl_lc}
\end{figure}

Our physics-based model is capable of reproducing the observed \textit{Kepler} light curve in detail
(the best-fitting solution, shown in Fig.~\ref{fig:mdl_lc}, has a reduced $ \chi^2 $ value of about 1.3).
The shape of the light curve is a natural result of the path and the size evolution of the dust grains
and does not require an occulting object consisting of multiple components
\citep[as suggested by][]{2014A&A...561A...3V}.
The low initial speed of the dust grains with respect to the planet
means that the angular density of extinction cross-section is very high at the head of the dust cloud,
while a large section of the tail has a more constant density (see also Fig.~\ref{fig:eom_example}).
Note that this finding is based on and relevant only to the phase-folded (i.e., averaged) light curve
and a tail with multiple components may still be necessary to explain the observed variability.

Also shown in Fig.~\ref{fig:mdl_lc} are 50 model realisations that were picked randomly from the chain,
to visualise the spread of the model solutions in data space.
We find that the chain contains many poorly fitting solutions,
with relatively shallow transits, prominent long egress tails, and other significant deviations from the observed profile.
In data space, these solutions occupy a large part of the region
allowed by the additional fitting requirements listed in Sect.~\ref{s:meth_lc_crit}.
This indicates that
the exploration of parameter space was not very efficient,
relying heavily on the additional fitting requirements,
with the $\chi^2$ statistic being of secondary importance.
As a result, the distribution in parameter space of model realisations in the Markov chain
may be a poor representation of the actual probability density distribution,
rather indicating the limits of the region of parameter space allowed by the observations.
Indeed, we find that good-fitting model solutions are spread throughout a large part of parameter space
explored by the MCMC algorithm
and do not cluster around the peak of
the distribution of model realisations.
For these reasons, we adopt a conservative approach in interpreting the MCMC results
and report $ 2 \sigma $ uncertainties in the remainder of this section,
which encompass most of the good-fitting models.

\subsection{Model-parameter constraints and correlations}
\label{s:res_constr}

\begin{figure*}
  \includegraphics[width=\linewidth]{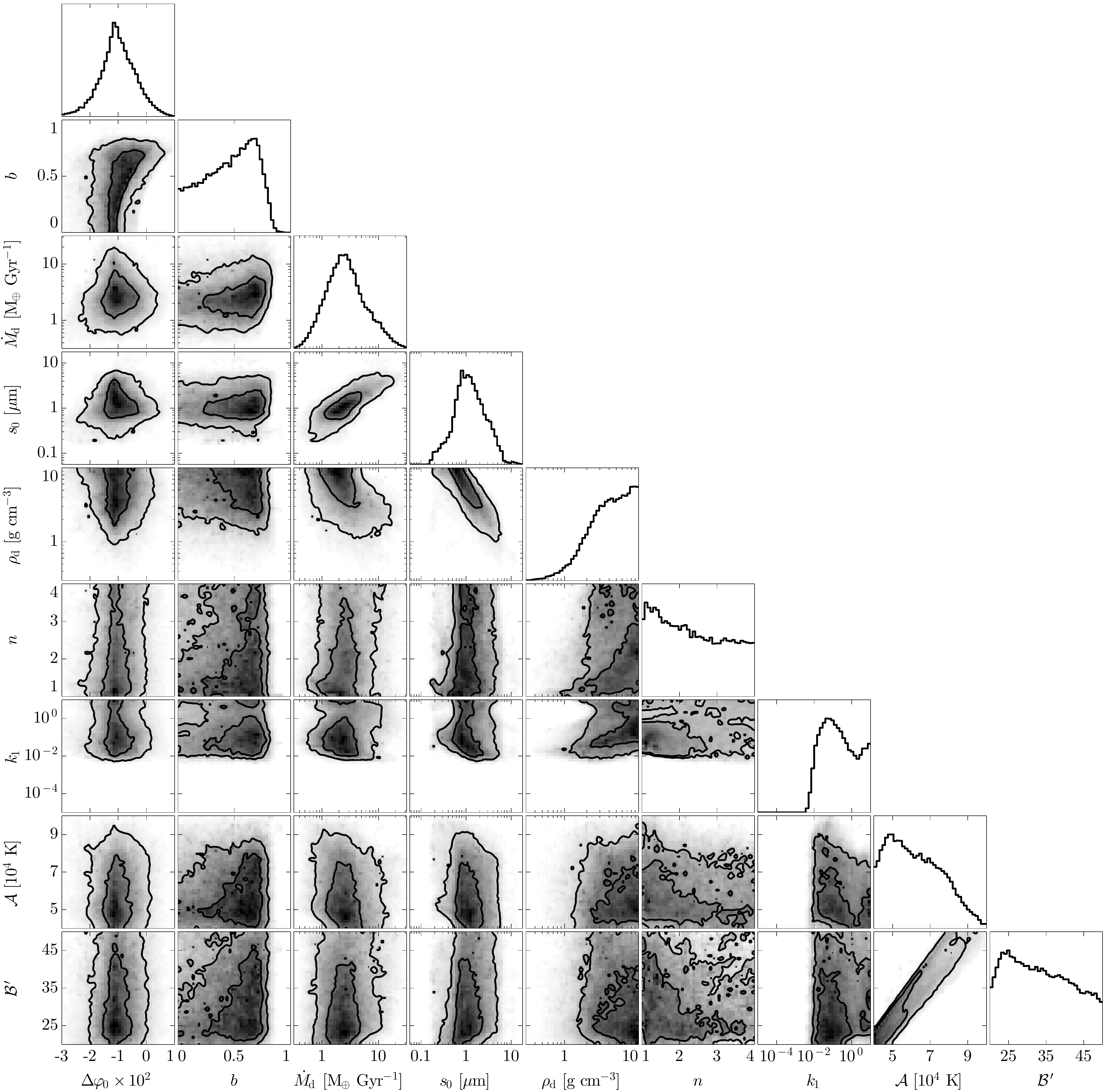}
  \caption{
  Posterior probability distribution over the free parameters of the model resulting from the MCMC analysis.
  See Table~\ref{tbl:free_pars} for a description of the free parameters.
  \textbf{Panels along the diagonal:}~One-dimensional projections (marginalised over all other parameters) of the probability density function.
  \textbf{Off-diagonal panels:}~Two-dimensional projections showing correlations between parameters.
  The grey-scale colours are proportional to the probability density (normalised per panel).
  Contours are drawn at $ 1 \sigma $ and $ 2 \sigma $ levels.
  }
  \label{fig:triangle}
\end{figure*}

The constrains on the model's free parameters resulting from the MCMC analysis are summarised in
Table~\ref{tbl:constr}.
Many of the individual free parameters are not well constrained.
By inspecting the result in more detail, however, we can extract useful constraints.
Figure~\ref{fig:triangle} gives a more extensive overview of the MCMC results,
showing one- and two-dimensional projections of the nine-dimensional posterior probability density function.
These indicate (1D) how symmetric or skewed the constraints on individual parameters are
and (2D) how pairs of free parameters are correlated.
Some three-dimensional projections of the probability density function were also inspected,
but these are not shown here.
We now discuss the correlations that occur between the free parameters of the model,
as well as some specifics of the constraints on individual parameters.

\begin{table}
  \centering
  \small
  \caption{Results of the MCMC analysis}
  \label{tbl:constr}
  {
  \renewcommand{\arraystretch}{1.15}
  \begin{tabular}{ccc}
  \hline
  \hline
  Parameter & Constraint ($ 2 \sigma $) & Notes \\
  \hline
  $ \Delta\varphi\sub{0} $ & $ -0.010 \substack{ +0.013 \\ -0.014 } $ & \\
  $ b $ & $ < 0.80 $ & \\
  $ \dot{M}\sub{d} $ & $ 2.5 \substack{ +13.1 \\ -1.9 } $~M$\sub{\oplus}$~Gyr$^{-1}$ & \\
  $ s\sub{0} $ & $ 1.2 \substack{ +4.4 \\ -0.9 } $~$\upmu$m & (a) \\
  $ \rho\sub{d} $ & $ > 0.9 $~g~cm$^{-3} $ & (a) \\
  $ n $ & unconstrained & \\
  $ k_1 $ & $ > 7.9 \times 10^{-3} $ \\
  $ \mathcal{A} $ & see note & (b) \\
  $ \mathcal{B}' $ & see note & (b) \\
  \hline
  \end{tabular}
  \tablefoot{
  The values reported here are medians of the marginalised probability density distribution.
  The uncertainties, as well as the upper and lower limits, reflect the 2.3\textsuperscript{th} and 97.7\textsuperscript{th} percentiles. \\
  \tablefoottext{a}{
  Combining $ s\sub{0} $ and $ \rho\sub{d} $ gives a constraint
  on the initial radiation-pressure-to-gravity ratio of
  $ \beta_0 = 0.029 \substack{ +0.094 \\ -0.016 } $ ($ 2 \sigma $).
  } \\
  \tablefoottext{b}{
  Individually, the sublimation parameters $ \mathcal{A} $ and $ \mathcal{B}' $ are unconstrained,
  but they
  are highly correlated and
  can be combined to give the constraint
  $ \log_{10}( J\sub{1700\,K} / ($g~cm$^{-2}$~s$^{-1}) ) = -7.3 \substack{ +2.6 \\ -2.9 } $ ($ 2 \sigma $).
  }
  }
  }
\end{table}

While the planet's dust mass loss rate $ \dot{M}\sub{d} $ and the initial size of the dust grains $ s\sub{0} $ are both individually relatively well constrained,
there is also a clear correlation between them.
Larger grains are associated with higher mass loss rates.
The mass loss rate is mostly constrained by the depth of the transit.
Because larger grains constitute less cross-section per unit mass,
a higher mass loss rate is needed to acquire the same transit depth with larger dust grains.
Also, larger grains have scattering phase functions that are more sharply peaked in the forward direction
and consequently the scattered-light component of their light curves (see Fig.~\ref{fig:lc_example})
is greater than that of small grains.
Since the scattered-light component can partly counteract the extinction signal,
larger grain sizes require more particles, and hence more mass, to yield the observed transit depth.

A clear anticorrelation is seen between the initial grain size $ s\sub{0} $ and the material density of the dust $ \rho\sub{d} $.
This can be explained by requirements on the $ \beta $ ratio,
which is inversely proportional to both.
Very small grains need to be dense in order to remain
 close to the star, where sublimation is efficient.
Large grains need to have a sufficiently low material density that
they drift far enough away from the planet within the allowed time to produce the tail seen in the light curve.
We can quantify these constraints by computing $ \beta\sub{0} $
(i.e., the $ \beta $ ratio at the moment the grain is released)
using Eq.~\eqref{eq:beta} with $ s = s\sub{0} $ and assuming $ \bar{Q}\sub{pr} = 1 $
(which holds within at most a factor 2).
Lines of constant $ \beta\sub{0} $ go parallel to the \mbox{$ s\sub{0} $-$ \rho\sub{d} $} correlation,
and consequently the distribution of $ \beta\sub{0} $ is relatively narrow (see note (a) in Table~\ref{tbl:constr}).
It peaks below the maximum $ \beta $ ratio possible around KIC~1255b (see Fig.~3 of \citetalias{2014A&A...572A..76V}),
but above the value for which initial launch velocities of the grains become important
(see Eq.~(23) of \citetalias{2014A&A...572A..76V}, and Sect.~9.2 and Fig.~15 of \citealt{2015ApJ...812..112S}).

\begin{table*}
\centering
\small
\caption{Parameters of the dust species considered in this study}
\label{tbl:dust_pars}
{
\renewcommand{\arraystretch}{1.15}
\begin{tabular}{l|r@{.}l|r@{.}lr@{ $\pm$ }lc|r@{ $\pm$ }lr@{ $\pm$ }lr@{ $\pm$ }lc}
\hline
\hline
Dust species & \multicolumn{2}{c|}{Density} & \multicolumn{5}{c|}{Complex refractive index\tablefootmark{a}} & \multicolumn{7}{c}{Sublimation parameters\tablefootmark{h}} \\
 & \multicolumn{2}{c|}{$ \rho\sub{d} $} & \multicolumn{2}{c}{$ n $} & \multicolumn{2}{c}{$ \log_{10}( k_1 ) $} & Refs. & \multicolumn{2}{c}{$ \mathcal{A} $} & \multicolumn{2}{c}{$ \mathcal{B}' $} & \multicolumn{2}{c}{$ \log_{10}( J\sub{1700\,K} ) $} & Refs. \\
 & \multicolumn{2}{c|}{[g~cm$^{-3}$]} & \multicolumn{2}{c}{} & \multicolumn{2}{c}{} & & \multicolumn{2}{c}{[$10^4$~K]} & \multicolumn{2}{c}{} & \multicolumn{2}{c}{\;\,\,[cgs]} & \\
\hline
Iron (Fe) & 7 & 87 & 3 & 5\tablefootmark{b} & $ 0.50 $ & $ 0.50 $\tablefootmark{c} & O88 & 4.84 & 0.12 & 31.3 & 0.7 & $ -4.8 $ & $ 0.4 $ & F04 \\
Silicon monoxide (SiO) & 2 & 13 & 1 & 9 & \multicolumn{2}{c}{see note (d)} & P85, W13 & 4.95 & 0.14 & 31.2 & 1.0 & $ -5.1 $ & $ 0.6 $ & G13 \\
Cryst. fayalite (Fe$_2$SiO$_4$) & 4 & 39 & 1 & 8 & $ -2.25 $ & $ 0.25 $ & Z11, F01 & 6.04 & 0.11 & 38.1 & 0.7\tablefootmark{i} & $ -4.8 $ & $ 0.4 $ & N94 \\
Cryst. enstatite (MgSiO$_3$) & 3 & 20 & 1 & 6\tablefootmark{e} & $ -4.25 $ & $ 0.25 $\tablefootmark{e} & D95, J98 & 6.89 & 0.88 & 37.8 & 5.0\tablefootmark{i} & $ -7.1 $ & $ 3.1 $ & M88, K91 \\
Cryst. forsterite (Mg$_2$SiO$_4$) & 3 & 27 & 1 & 6 & $ -3.75 $ & $ 0.25 $ & Z11, F01 & 6.53 & 0.40 & 34.3 & 2.5 & $ -7.8 $ & $ 1.5 $ & G10, N94 \\
Quartz (SiO$_2$) & 2 & 60 & 1 & 6\tablefootmark{f} & $ -3.25 $ & $ 0.25 $\tablefootmark{f} & Z13 & 6.94 & 0.34 & 35.1 & 1.8 & $ -8.5 $ & $ 1.2 $ & H90 \\
Corundum (Al$_2$O$_3$) & 4 & 00 & 1 & 6\tablefootmark{g} & $ -1.75 $ & $ 0.25 $\tablefootmark{c,g} & K95 & 7.74 & 0.39\tablefootmark{j} & 39.3 & 2.0\tablefootmark{j} & $ -8.7 $ & $ 1.3 $ & S04, L08 \\
Silicon carbide (SiC) & 3 & 22 & 2 & 5 & $ -3.50 $ & $ 0.25 $ & La93 & 7.85 & 0.39\tablefootmark{j} & 37.4 & 1.9\tablefootmark{i,j} & $ -9.8 $ & $ 1.3 $ & Li93 \\
Graphite (C) & ~~~2 & 16 & 3 & 2\tablefootmark{b} & $ -0.25 $ & $ 0.50 $ & D84 & 9.36 & 0.05 & 36.2 & 1.8\tablefootmark{i,j} & $ -14.2 $ & $ 0.8 $ & Z73 \\
\hline
\end{tabular}
\tablefoot{The list only includes crystalline species
because amorphous dust is expected to anneal rapidly \citep[see Sect.~4.2.3 of][]{2002Icar..159..529K}. \\
\tablefoottext{a}{See text and Appendix~\ref{app:lnk_recipe} on how the values for $ n $ and $ k_1 $ were derived
from the wavelength-dependent complex-refractive-index data.} \\
\tablefoottext{b}{For iron and graphite, $ n ( \lambda ) $ steadily rises in the wavelength regime considered
(see Fig.~\ref{fig:lnk}).} \\
\tablefoottext{c}{For iron and corundum,
the best-matching $ k_1 $ values
give temperatures that are slightly too low (see Appendix~\ref{app:lnk_recipe}).} \\
\tablefoottext{d}{Our two-parameter complex-refractive-index recipe cannot reproduce
the $ T\sub{d} ( s ) $ profile of SiO (see Appendix~\ref{app:lnk_recipe}).} \\
\tablefoottext{e}{The optical properties at wavelengths below 8~$\upmu$m use amorphous enstatite.} \\
\tablefoottext{f}{The $ n ( \lambda ) $ and $ k ( \lambda ) $ data we use for quartz only cover wavelengths of 3~$\upmu$m and higher.
To compute $ n $ and $ k_1 $, these were extrapolated down.} \\
\tablefoottext{g}{The data of \citet{1995Icar..114..203K} were obtained using a material consisting mostly of $\upgamma$-Al$_2$O$_3$,
while corundum is $\upalpha$-Al$_2$O$_3$.} \\
\tablefoottext{h}{Additional notes and details on the sublimation parameters can be found in Table~3 of \citetalias{2014A&A...572A..76V}.} \\
\tablefoottext{i}{For these materials, no measurements of the evaporation coefficient $ \alpha $ are available.
In the computation of $ \mathcal{B}' $, we arbitrarily adopt $ \alpha = 0.1 $.} \\
\tablefoottext{j}{For sublimation parameters without a reported uncertainty,
we set the standard deviation on $ \mathcal{A} $ and/or $ \mathcal{B} $ to 5\%,
which is comparable to the level of uncertainty of sublimation parameters that do include an error bar.}
}
 \tablebib{
 D84~\citet{1984ApJ...285...89D};
 D95~\citet{1995A&A...300..503D};
 F01~\citet{2001A&A...378..228F};
 F04~\citet{ferguson2004};
 G10~\citet{2010LNP...815...61G};
 G13~\citet{2013A&A...555A.119G};
 H90~\citet{1990Natur.347...53H};
 J98~\citet{1998A&A...339..904J};
 K91~\citet{kushiro91};
 K95~\citet{1995Icar..114..203K};
 La93~\citet{1993ApJ...402..441L};
 Li93~\citet{Lilov199365};
 L08~\citet{Lihrmann2008649};
 M88~\citet{1988AmMin..73....1M};
 N94~\citet{1994GeCoA..58.1951N};
 O88~\citet{1988ApOpt..27.1203O};
 P85~\citet{1985hocs.book.....P};
 S04~\citet{2004Icar..169..216S};
 W13~\citet{2013A&A...553A..92W};
 Z73~\citet{1973JChPh..59.2966Z}.
 Z11~\citet{2011A&A...526A..68Z};
 Z13~\citet{2013A&A...553A..81Z}.
 }
}
\end{table*}

While the probability density distribution found by the MCMC analysis peaks at
initial grain sizes of about $ 1 $~$\upmu$m
and mass loss rates of a few M$\sub{\oplus}$~Gyr$^{-1}$,
good fits are found over a broad range of values for these parameters.
Furthermore, the detailed shapes of different parts of the transit light curve
are best reproduced by different parts of the free-parameter space.
In particular, the best results regarding the pre-ingress brightening
(which is sensitive to the grain size because that determines the shape of the scattering phase function)
are achieved
with initial grain sizes of 0.2 to 0.3~$\upmu$m.
Larger grains provide too much forward scattering
and smaller sizes (if allowed) give light curves that are too flat.
Because of the correlations between $ \dot{M}\sub{d} $, $ s\sub{0} $, and $ \rho\sub{d} $ described above,
allowed model solutions with
these initial grain sizes correspond to
dust mass loss rates of the order of 1~M$\sub{\oplus}$~Gyr$^{-1}$
and material densities close to the edge of the parameter space we consider (10~g~cm$^{-3}$).

The optical properties of the dust material influence the constraints mostly through their effects on the dust temperatures.
For $ k_1 $ (the imaginary part of the complex refractive index for $ \lambda < 8 $~$\upmu$m),
there is a sharp lower limit just below $ k_1 = 10^{-2} $.
This is an effect of how the
$ T\sub{d} ( s ) $ profile
changes with $ k_1 $.
For low values of $ k_1 $, the dust temperature only goes down with decreasing grain size (see Fig.~\ref{fig:temp_comp}).
For dust with such a profile, the sublimation rate quickly goes down as the grain becomes smaller
and as a result the grains do not sublimate.
This makes it impossible to create a tail of finite length with a \mbox{low-$ k_1 $} material.
The real part of the complex refractive index $ n $ has a weaker effect on dust temperatures and hence it is not constrained,
although there is a small preference for lower values.
This may be caused by the smoother $ T\sub{d} ( s ) $ profiles associated with lower $ n $ values (see Fig.~\ref{fig:temp_comp}).

Finally, the sublimation parameters $ \mathcal{A} $ and $ \mathcal{B}' $ are unconstrained individually,\footnote{%
The lack of solutions with high $ \mathcal{A} $ is an effect of the bound on $ \mathcal{B}' $.}
but there is a clear correlation between them.
This is to be expected, since
they both appear in the exponent in Eq.~\eqref{eq:pres_vap}
and the relative range in dust temperature reached by the grains is small.
To aid our further analysis, we define the new parameter
\begin{equation}
  \label{eq:j1700}
  J\sub{1700\,K} \equiv J( T\sub{d} = 1700~\mathrm{K} ),
\end{equation}
i.e., the mass loss flux (units: [g~cm$^{-2}$~s$^{-1}$]; positive for mass loss) from a dust grain at a temperature of 1700~K,
with $ J $ defined in Eq.~\eqref{eq:dsdt}.
The value of $ T\sub{d} = 1700~\mathrm{K} $ was chosen to minimise the width of the $ J\sub{1700\,K} $ distribution
(i.e., lines of constant $ J\sub{1700\,K} $ are parallel to the $ \mathcal{A} $-$ \mathcal{B}' $ correlation)
and can be thought of
as the typical temperature of dust grains in well-fitting model realisations.
However, the distribution we find for $ J\sub{1700\,K} $ is still relatively broad (see note (b) in Table~\ref{tbl:constr}),
reflecting both the range in sublimation rates that give good-fitting light curves
and the range in temperature reachable by combining different optical properties and grain sizes.
Besides the correlation,
there is also a small preference for the bottom left corner of the \mbox{$ \mathcal{A} $-$ \mathcal{B}' $} diagram,
which corresponds to lower temperature sensitivities.

\subsection{Comparison with laboratory-measured dust properties}
\label{s:res_species}

\begin{figure*}
  \includegraphics[width=\linewidth]{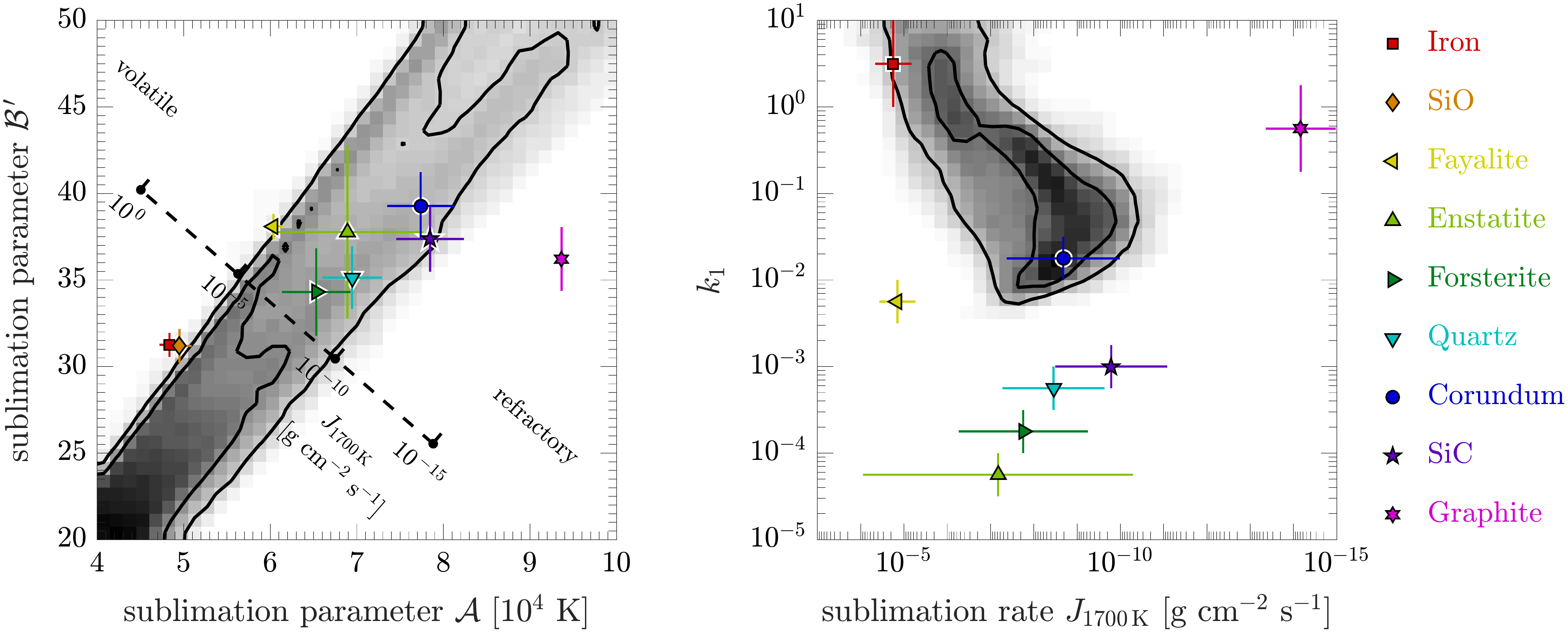}
  \caption{
  Comparison between our model results and the laboratory-measured properties of some real dust species,
  as listed in Table~\ref{tbl:dust_pars},
  for the model parameters that provide meaningful constraints on the dust composition.
  \textbf{Left:}~Constraints on the parameters that describe the temperature dependence of dust sublimation.
  The dashed diagonal axis shows how
  $ J\sub{1700\,K} $ (the sublimation rate of dust at 1700~K)
  varies with $ \mathcal{A} $ and  $ \mathcal{B}' $.
  \textbf{Right:}~Combined constraints on $ k_1 $ (the imaginary part of the complex refractive index for $ \lambda < 8 $~$\upmu$m)
  and the 1700~K sublimation rate.
  Note that the sublimation rate decreases to the right.
  }
  \label{fig:heatmaps}
\end{figure*}

The main purpose of this work is to understand how well the composition of the dust in the tail of KIC~1255b
can be constrained by modelling its light curve.
To see how the outcome of the MCMC analysis constrains the properties of the dust grains,
we now compare the posterior probability density function
in the five composition parameters ($ \rho\sub{d} $, $ n $, $ k_1 $, $ \mathcal{A} $, and $ \mathcal{B}' $)
to the values of these parameters for nine selected dust species.

Table~\ref{tbl:dust_pars} lists the values of the composition parameters for the materials we want to test.
The method of deriving values of $ n $ and $ k_1 $ for each dust species from their wavelength-dependent complex-refractive-index data
is described in Appendix~\ref{app:lnk_recipe}.
In short, the values listed for $ n $ are valid for the wavelength range 0.4 to 4~$\upmu$m,
which covers the \textit{Kepler} bandpass as well as the peak of the stellar spectrum.
For $ k_1 $, we take the values that give the best match between the $ T\sub{d} ( s ) $ profiles
computed using the full wavelength-dependent complex refractive indices of the various materials
and using our simple two-parameter prescription.
Regarding the sublimation parameters, further details and notes can be found in Table~3 of \citetalias{2014A&A...572A..76V}.
This includes the values of $ \alpha $, $ \mu $, and $ \mathcal{B} $ used in the calculation of $ \mathcal{B}' $.

Of the five free parameters of our model that describe the dust properties,
two cannot be used to exclude species from Table~\ref{tbl:dust_pars}:
the real part of the complex refractive index $ n $ is not constrained by the data
and the lower-limit requirement for the material density $ \rho\sub{d} $ is met by all dust species under consideration.
The three remaining composition parameters ($ k_1 $, $ \mathcal{A} $, and $ \mathcal{B}' $)
do give meaningful constraints.
Figure~\ref{fig:heatmaps} shows the marginalised probability density distribution for these parameters,
together with the values of the materials from Table~\ref{tbl:dust_pars}.
The two panels are essentially two different projections of the same three-dimensional confidence region,
with $ J\sub{1700\,K} $ being used to define an axis perpendicular to the \mbox{$ \mathcal{A} $-$ \mathcal{B}' $} correlation.

Comparison of the model results with the parameters of the dust species reveals
that many of the tested materials cannot reproduce the observed transit profile.
Only corundum (i.e., crystalline aluminium oxide) gives a satisfactory fit.
We now briefly discuss each of the materials individually.
\begin{itemize}
\item
\textit{Iron} gives a reasonable fit in the \mbox{$ J\sub{1700\,K} $-$ k_1 $} projection.
However, its sublimation parameters, which are established to higher accuracy than any of the other materials we consider,
indicate that it is too volatile.
Taking into account that its temperatures are slightly higher than predicted by its $ k_1 $ value (see Appendix~\ref{app:lnk_recipe}),
which would increase the distance between the iron symbol and the edge of the allowed region of parameter space,
we deem grains of pure iron unlikely.
\item
\textit{Silicon monoxide} is not shown in the right panel of Fig.~\ref{fig:heatmaps},
because its $ T\sub{d} ( s ) $ profile cannot be reproduced using our complex-refractive-index recipe for any value of $ k_1 $ (see Appendix~\ref{app:lnk_recipe}).
Since it reaches temperatures much higher than any that can be produced by the recipe,
and its sublimation parameters are on the ``volatile edge'' of the allowed region,
this material can be excluded as too hot and volatile.
\item
For \textit{fayalite} (the iron-rich end-member of olivine), both the sublimation rate and $ k_1 $ individually give a marginal fit,
but the combined constraints disfavour this material.
\item
The silicates \textit{enstatite} (the magnesium-rich end-member of pyroxene),
\textit{forsterite} (the magnesium-rich end-member of olivine), and \textit{quartz}
all have sublimation parameters that are consistent with the model constraints,
but their $ k_1 $ values are too low
(i.e., in pure form they are too transparent at visible and near-infrared wavelengths).
\item
Of the tested materials,
\textit{corundum} is the only one
with properties that lie completely within the allowed region of parameter space.
It should be noted that, as for iron, our complex-refractive-index recipe gives temperatures for corundum that are slightly lower
than the $ T\sub{d} ( s ) $ profile computed using the full wavelength-dependent complex refractive indices (see Appendix~\ref{app:lnk_recipe}).
This discrepancy means that corundum sublimates somewhat faster than its position in \mbox{$ J\sub{1700\,K} $-$ k_1 $} space suggests.
However, since corundum is positioned farther from the ``volatile edge'' of the allowed region of parameter space,
the argument used to disfavour iron and silicon monoxide does not apply to corundum.
\item
For \textit{silicon carbide}, the sublimation rate is marginally compatible with the constraints,
but the value of $ k_1 $ is too low.
\item
Pure \textit{graphite} grains can clearly be excluded, since they are too refractory.
\end{itemize}

Of course, Table~\ref{tbl:dust_pars} is a very incomplete list of possible dust species,
in particular because it only includes pure materials.\footnote{%
An important limiting factor in compiling a list of possible dust species for this research
is the lack of reliable and consistent laboratory measurements of sublimation properties.
Especially measurements of the accommodation coefficient $ \alpha $ are sparse.}
Because of the parametric description we use for the dust material,
however,
the results of the MCMC analysis
provide constraints on the material properties that are independent of 
the availability of laboratory measurements
and can be used to test any material for which good laboratory measurements become available.


\section{Discussion}
\label{s:disc}

\subsection{Comparison with previous findings}
\label{s:disc_compare}

We reach similar conclusions about the dust composition of KIC~1255b as in \citetalias{2014A&A...572A..76V}.
In particular, both analyses find that corundum is the only material out of the nine tested species that can yield the right tail length.
Previously, KIC~1255b was suggested to exhibit dust composed of pyroxene \citep{2012ApJ...752....1R},
based on sublimation times derived by \citet{2002Icar..159..529K}.
The discrepancy can be traced back to a difference in complex refractive indices and sublimation parameters used.
We use up-to-date laboratory data (see Table~\ref{tbl:dust_pars}),
and note that for pyroxene \citet{2002Icar..159..529K} use sublimation parameters measured for SiO$_2$.

Aluminium has a relatively low cosmic abundance and it is probably only a minor constituent of rocky planets
\citep[e.g., Table~2 of][]{2014AREPS..42...45J}.
Therefore, a tail consisting of corundum dust is surprising,
especially in combination with the high mass loss rate inferred for KIC~1255b.
One possible explanation for this apparent discrepancy is that conditions in the planet's atmosphere may favour the condensation of a particular dust species.
This would be the case, for example, if the gas outflow becomes too tenuous for the condensation of dust while its temperature is still high.
Due to its high condensation temperature, corundum is one of the first species to condense out of a mix of gasses.
Another possible explanation for a corundum-dust tail is the distillation of a lava ocean on the surface of the planet
to the point that the residue consists mostly of calcium and aluminium oxides
(see Sect.~4.3 of \citetalias{2014A&A...572A..76V} and Sect.~6 of \citealt{2011Icar..213....1L}).

The dust mass loss rates that we find are consistent with those derived analytically in \citetalias{2014A&A...572A..76V},
but our constraints extend to higher values than earlier order-of-magnitude estimates
\citep[0.1 to 1~M$\sub{\oplus}$~Gyr$^{-1}$,][]{2012ApJ...752....1R,2013MNRAS.433.2294P,2013ApJ...776L...6K}.
The reason for this difference is that
we take into account the scattering of star light by dust particles.
Forward scattering can fill up part of the transit light curve,
leading to higher mass loss rates to achieve the same transit depth.

Based on the relation between a planet's mass and its mass loss rate found by \citet{2013MNRAS.433.2294P},
our results imply a relatively low present-day planetary mass
and consequently a short lifetime remaining until the complete destruction of the planet.
Also, the highest mass loss rates we find are close to the free streaming limit.
Note, however, that mass loss rates in the model of \citet{2013MNRAS.433.2294P} depend sensitively
on the assumed composition of the atmosphere and the dust (see their Figs.~2 and 10),
because this sets the (boundary-condition) gas density at the planetary surface via the composition-dependent vapour pressure.

\subsection{An optically thick coma?}
\label{s:disc_optical_depth}

\begin{figure}
  \includegraphics[width=\linewidth]{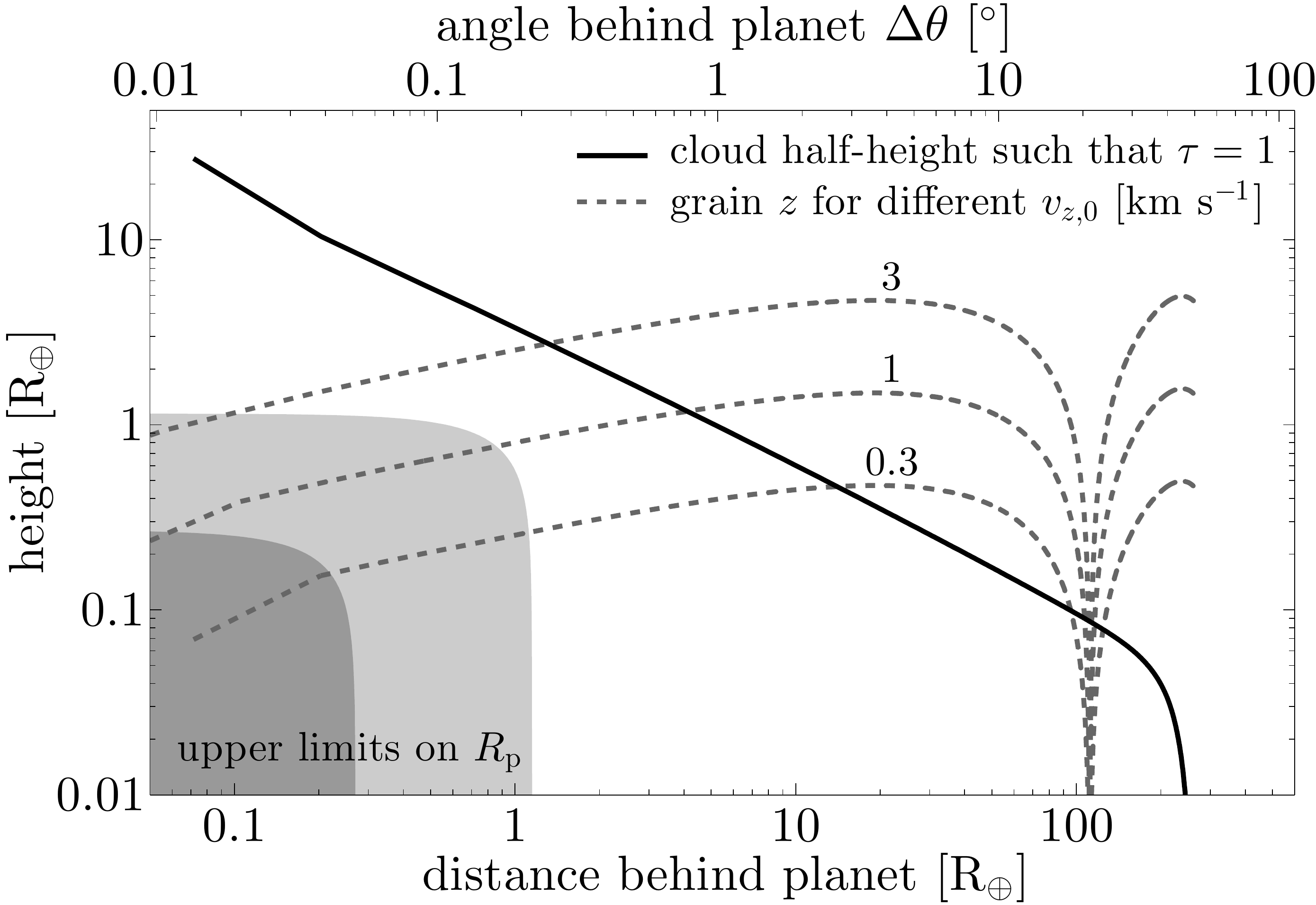}
  \caption{
  Diagram to illustrate for what part of the tail the assumption of an optically thin dust cloud is valid.
  The solid line shows $ h_{ \tau = 1 } $,
  the dust cloud half-height for which the radial optical depth equals unity.
  The dashed lines give the heights above the planet's orbital plane
  reached by dust grains that are launched vertically from the planet
  with different initial vertical speeds $ v_{z,0} $.
  Distances on the bottom axis are measured along the planet's orbit.
  All curves are drawn for a typical model realisation.
  The shaded regions in the lower left indicate
  upper limits on the planet's radius based on
  (1)~the absence of transits in part of the \textit{Kepler} data
  \citep[light grey; $ R\sub{p} < 1.15 $~R$\sub{\oplus} \; ( 1 \sigma ) $;][]{2012A&A...545L...5B}
  and (2)~the planet-mass-dependent mass loss rate
  \citep[dark grey; $ R\sub{p} \lesssim 0.3 $~R$\sub{\oplus}$, model-dependent;][]{2013MNRAS.433.2294P}.
  }
  \label{fig:opt_depth_chk}
\end{figure}

Throughout this work, as well as in \citetalias{2014A&A...572A..76V},
we have assumed that the dust tail is optically thin in the radial direction.
To check the validity of this assumption,
we compute the height with respect to the planet's orbital plane
that the dust tail needs to extend in order to
have a radial optical depth of unity.
We denote half of this height with $ h_{ \tau = 1 } $
and  in Fig.~\ref{fig:opt_depth_chk} this quantity is shown for a typical model realisation.
It is compared with the height that the cloud can be expected to have
based on the maximum possible size of the source region (i.e., upper limits on the size of the planet),
and the likely vertical spreading caused by non-zero launch speeds in the out-of-the-plane direction.
For example, at a distance of 10~R$ \sub{\oplus} $ behind the planet,
the tail should extend more than 0.6~R$ \sub{\oplus} $ above the planet's orbital plane in order to be radially optically thin.
This vertical extent is reached if the source region itself is larger than 0.6~R$ \sub{\oplus} $
and/or if particles are launched from the planet with vertical launch speeds of about 0.5\,km\,s$^{-1}$ or higher.

Speeds of $ \sim $1~km~s$^{-1}$ are reached in the thermal wind
that launches the dust particles out of the planet's atmosphere \citep[see Fig.~3 of][]{2013MNRAS.433.2294P}.
We do not expect launch speeds greater than a few km~s$^{-1}$,
since these would result in dust dynamics controlled by their launch velocities, rather than radiation pressure
(see Eq.~(23) of \citetalias{2014A&A...572A..76V}, together with the constraint on $ \beta\sub{0} $ from Table~\ref{tbl:constr})
and (assuming the dust is emitted more or less isotropically) a transit profile showing a dust streamer leading the planet as well as a tail,
contrary to what is observed.

Assuming vertical launch speeds of $ \sim $1~km~s$^{-1}$, Fig.~\ref{fig:opt_depth_chk} shows that
the first few R$ \sub{\oplus} $ of the dust cloud may be optically thick.
Beyond a distance of $ \sim $10~R$ \sub{\oplus} $ behind the planet,
we expect the dust tail to become optically thin.
In summary, the optical-depth assumption is valid for most of the dusty tail,
but possibly not for the first few R$ \sub{\oplus} $.
This inaccuracy, however, may be mitigated
by the large solid angle of the star at the distance of the dust cloud.
That is, dust grains that are obscured by others in the radial direction
can still be reached by radiation emanating from higher stellar latitudes,
in particular if the dust cloud is more extended in the radial direction than in the vertical, which is likely.

The region up to 10~R$ \sub{\oplus} $ is responsible for about half of the transit depth as computed under the assumption that it is optically thin.
Therefore, optical-depth effects could have a significant impact on the shape of the light curve,
as well as the wavelength dependence of the transit depth.
However, it is important to note that the optical depth in the radial direction is not the same as the optical depth from the star to the observer.
The inclination of the planet's orbital plane with respect to the line of sight
means that the optical depth towards the observer will be less than the radial optical depth.
This can be a strong effect, since dust motion in the planet's orbital plane are substantial,
and, for an inclined orbit, they contribute to the vertical extent of the dust cloud with respect to the line of sight.
Nevertheless, the high radial optical depth close to the planet warrants further investigation
since it will affect the radiation pressure on the dust grains as well as their temperatures and sublimation rates.

More generally, dust particles in the close vicinity of the planet will be affected by
the planet's gravity and interactions with the outflowing gas,
processes that were ignored in our modelling.
Although the region where these processes are important
(within about 1~R$ \sub{\oplus} $ of the planet)
is probably smaller than the tentative optically thick coma,
it is difficult to draw firm conclusions
about the behaviour of the dust tail very close to the planet
without a more comprehensive model.

Finally, the dashed curves in Fig.~\ref{fig:opt_depth_chk} show minima just beyond 100~R$ \sub{\oplus} $,
which occur because 
dust particles on inclined orbits cross the orbital plane of the planet halfway through their own orbit.
In principle, this could give
another possible explanation for the small decrement in flux tentatively detected
in the egress of the \textit{Kepler} short cadence light curve of KIC~1255b \citep{2014ApJ...786..100C}.
In this scenario, mutual shadowing (hence, an increase in optical depth) at the knot halfway along the orbit of the dust grains
would cause a temporary brightening in the egress and an apparent dimming afterwards, when the shadowing becomes insignificant again.
However, to reach an optical depth higher than unity,
the source area would have to be significantly smaller than
the required $ h_{ \tau = 1 } $ at the distance of the minimum,
which is less than 0.1~$ R\sub{\oplus} $.
Furthermore, small differences in initial grain size
(and therefore in $ \beta $)
as well as in initial launch speed and direction can easily wash out this optical-depth effect.


\section{Conclusions}
\label{s:conclusions}

We have developed a numerical model to simulate the dusty tails of evaporating planets and their transit light curves,
with the primarily goal of putting constraints on the composition of the dust in such tails.
The model solves the dynamics and sublimation of dust particles in the orbital plane of the planet (2D)
and then generates a synthetic light curve of the dust cloud transiting the star
(after rotating the dust cloud to take into account its inclined orbit; hence, 3D),
which can be compared with an observed light curve.
We applied this model to the phase-folded \textit{Kepler} light curve of the prototypical evaporating planet KIC~1255b,
using an MCMC optimisation technique to constrain the free parameters of the model,
including those describing the dust composition.

Although the precise best-fit values and uncertainties we find for the model parameters
may depend on modelling details (e.g., simulating only a single initial grain size),
our analysis shows that
by using a physically motivated model
it is possible to put meaningful constraints on the composition of the dust in the tail of an evaporating planet
based on the shape of its broadband transit light curve.
Since the dust composition is related to that of the planet,
such constraints can provide helpful input for theories of planet formation and evolution.

Regarding KIC~1255b, we draw the following conclusions.
\begin{enumerate}
\item
\textit{Dust composition.}
We find that only certain combinations of material properties
(specifically, of sublimation parameters and the imaginary part of the complex refractive index)
can reproduce the observed transit profile.
To obtain the observed tail length while avoiding the correlations between subsequent transits
that arise when grains survive longer than an orbital period of the planet,
the dust grains need to have the right sublimation rate and temperature.
The constraints we find allow us to rule out or disfavour many of the pure materials we tested for the dust composition (see Fig.~\ref{fig:heatmaps}):
iron, silicon monoxide, fayalite, enstatite, forsterite, quartz, silicon carbide, and graphite.
The only material we found to match the constraints is corundum (i.e., crystalline aluminium oxide).
Grains made of combinations of the tested materials, however, cannot be ruled out.
The present results agree with those found earlier using a semi-analytical approach \citepalias{2014A&A...572A..76V},
which gives credence to this simpler method.
\item
\textit{Grain sizes.}
We simulate the dust cloud assuming the dust grains all have the same initial size,
but let this size evolve as a result of sublimation.
Good fits to the observed light curve are produced by initial grain sizes between $ 0.2 $ and $ 5.6 $~$\upmu$m ($ 2 \sigma $ range).
The shape of the pre-ingress brightening favours initial grain sizes
at the lower edge of this range (0.2 to 0.3~$\upmu$m).
\item
\textit{Mass loss rate.}
We find that the planet loses
0.6 to 15.6~M$\sub{\oplus}$~Gyr$^{-1}$
in dust alone ($ 2 \sigma $ range).
\item
\textit{Tail morphology.}
It is not necessary to invoke an object consisting of multiple components (e.g., a coma and a tail)
to explain the detailed shape of the averaged transit light curve.
The shape emerges naturally from
the distribution of the dust extinction cross-section in the tail
(see Fig.~\ref{fig:eom_example}).
We also find evidence
that the head of the dust cloud may be optically thick in the radial direction (see Fig.~\ref{fig:opt_depth_chk}).
\end{enumerate}

\begin{acknowledgements}
We thank the anonymous referee for a thorough review of the manuscript.
Support for this work was provided by NASA
through Hubble Fellowship grant HST-HF2-51336 awarded by the Space
Telescope Science Institute, which is operated by the Association of
Universities for Research in Astronomy, Inc., for NASA, under contract
NAS5-26555.
The research leading to the presented results has received funding from the European Research Council
under the European Community's Seventh Framework Programme (FP7/2007-2013) / ERC grant agreement no 338251 (StellarAges).
\end{acknowledgements}


\bibliographystyle{aa}
\bibliography{../../../latex/bibtex/bib_ads.bib}

\clearpage

\begin{appendix}

\section{Evolutionary status of KIC~12557548}
\label{app:star_param}

\begin{table*}
  \centering
  \small
  \caption{Observational estimates of the stellar parameters of KIC~12557548}
  \label{tbl:kic_pars}
  {
  \renewcommand{\arraystretch}{1.2}
  \begin{tabular}{lr@{ $\pm$ }lr@{ $\pm$ }lr@{ $\pm$ }lcc}
  \hline
  \hline
  Reference & \multicolumn{2}{c}{$ T\sub{eff,\star} $ [K]} & \multicolumn{2}{c}{$ \log \varg $ [cgs]} & \multicolumn{2}{c}{[Fe$/$H]} & Evolutionary status & Method \\
  \hline
  Kepler Input Catalogue\tablefootmark{a} & $ 4400 $ & $ 200 $ & $ 4.6 $ & $ 0.5 $ & $ -0.2 $ & $ 0.5 $ & main-sequence star & photometry \\
  \citet{2012ApJ...752....1R} & $ 4300 $ & $ 250 $ & \multicolumn{2}{c}{} & \multicolumn{2}{c}{} & main-sequence star & low-resolution spectroscopy \\
  \citet{2013ApJ...776L...6K} & $ 4950 $ & $ 70 $ & $ 3.9 $ & $ 0.2 $ & $ 0.09 $ & $ 0.09 $ & subgiant & high-resolution spectroscopy \\
  \citet{2014ApJS..211....2H}\tablefootmark{b} & \multicolumn{2}{l}{$ 4550 \; \substack{+140 \\ -131} $} & \multicolumn{2}{l}{$ 4.622\substack{+0.043 \\ -0.036} $} & $ -0.180 $ & $ 0.320 $ & main-sequence star & photometry \\
  \hline
  \end{tabular}
  \tablefoot{
  From left to right: effective temperature, surface gravity, and metallicity.
  \tablefoottext{a}{\citet{2011AJ....142..112B}.}
  \tablefoottext{b}{The estimate of \citet{2014ApJS..211....2H} incorporates priors (see their Sect.~2).
  This explains their improved uncertainties with respect to the Kepler Input Catalogue, especially on $ \log \varg $.}
  }
  }
\end{table*}

The literature contains diverging observational estimates of the stellar parameters of KIC~12557548,
casting uncertainty on its evolutionary status (see Table~\ref{tbl:kic_pars}).
The star is generally found to be a \mbox{mid-K} dwarf,
but \citet{2013ApJ...776L...6K} report stellar parameters that point to a subgiant status.
If the host star is indeed evolved off the main sequence,
this has important consequences for the story of the evaporating planet.
It could mean that the evaporation of the planet was triggered by the evolution of the star:
i.e., that the planet was stable at its present orbital distance throughout the main-sequence lifetime of the star,
and has only recently started losing (substantial amounts of) mass due to an increase in stellar irradiation.
Therefore, it is important to have a good understanding of the evolutionary status of KIC~12557548.
In addition, the constraints derived from our dust tail modelling depend on the properties we assume for the host star.
This appendix describes the checks we did
to distinguish between the two scenarios (main-sequence star vs. subgiant).

We first performed an astroseismic analysis,
which entails searching for evidence of solar-like oscillations in the light curve of KIC~12557548.
The frequencies of these oscillations scale with stellar properties,
and the method has the potential to pin down a star's evolutionary status with high accuracy
\citep{2013ARA&A..51..353C}.
Specifically, we inspected
the power-spectrum of the decorrelated and detrended long-cadence \textit{Kepler} light curve
with the transit signal (phases $ \varphi \in [-0.15, 0.2] $) masked out.
Unfortunately, no significant solar-like oscillations could be detected.
In principle, their non-detection can be used to obtain
a lower limit on the star's surface gravity \citep{2014ApJ...783..123C}.
Given the target's faintness, however,
we expect the \textit{Kepler} data to be of insufficient quality to yield a meaningful constraint.

\begin{figure}
  \includegraphics[width=\linewidth]{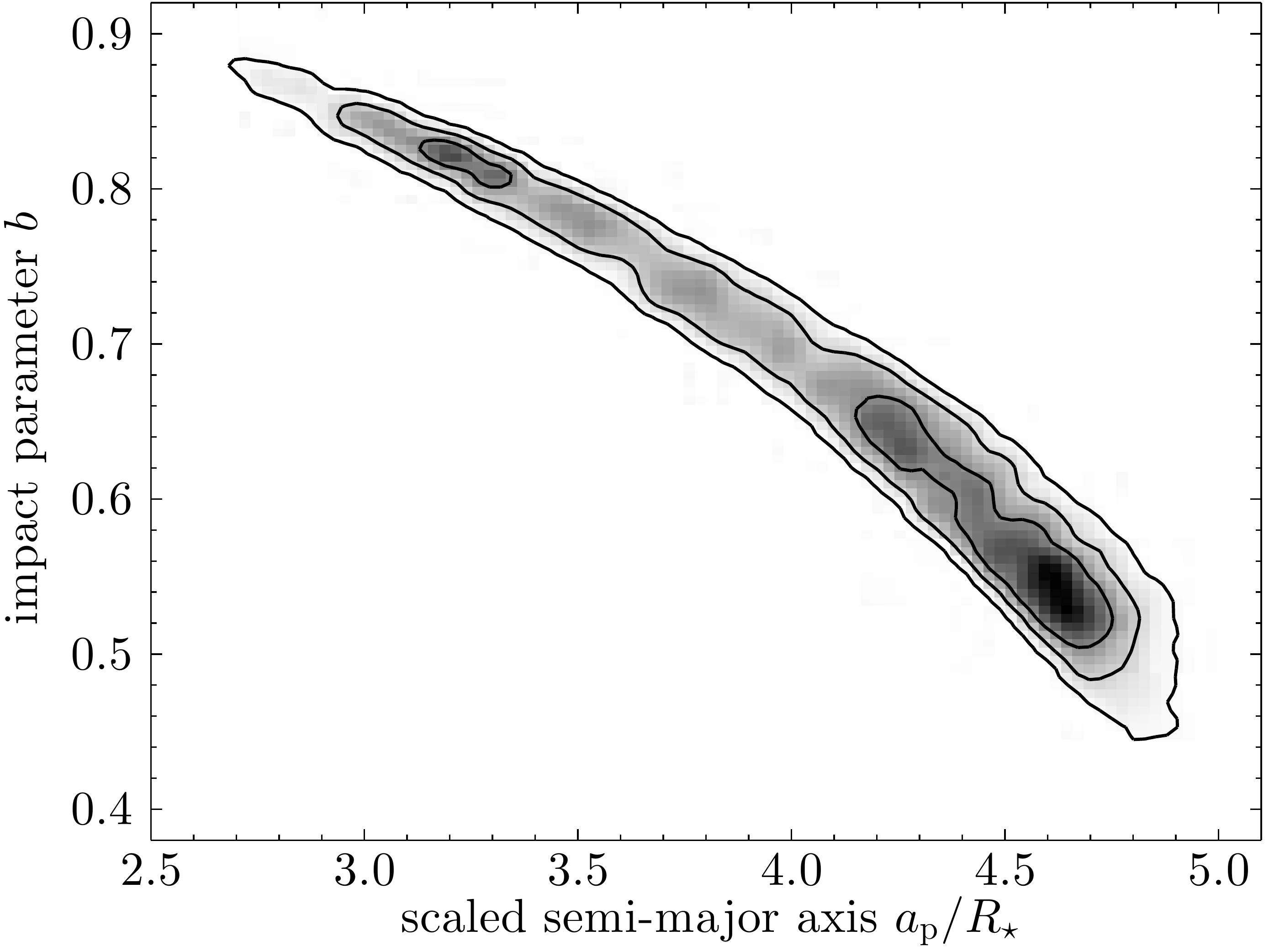}
  \caption{
  Combined constraints on the scaled semi-major axis $ a\sub{p} / R\sub{\star} $ and impact parameter $ b $ of KIC~1255b
  from the \mbox{1-D} dust cloud model of \citet{2012A&A...545L...5B}
  with $ a\sub{p} / R\sub{\star} $ added as an extra free parameter.
  The contours mark the $1\upsigma$, $2\upsigma$, and $3\upsigma$ confidence regions.
  }
  \label{fig:a_vs_b}
\end{figure}

Another way to constrain the properties of KIC~12557548 is to use the transit signal,
which contains information about the star's mean density.
Using Kepler's third law,
it is possible to express a star's mean density $ \rho\sub{\star} $
in terms of the orbital period $ P\sub{p} $ and
scaled semi-major axis $ a\sub{p} / R\sub{\star} $ of an orbiting planet:
\begin{equation}
  \label{eq:rho_star}
  \rho\sub{\star} = \frac{ 3 \pi }{ G P\sub{p}^2 } \left( \frac{ a\sub{p} }{ R\sub{\star} } \right)^3.
\end{equation}
For a spherical transiting planet,
$ a\sub{p} / R\sub{\star} $ can be computed analytically from the shape of the light curve \citep{2003ApJ...585.1038S}.
It is mainly sensitive to the ratio of transit duration to orbital period, modulated by the impact parameter $ b $
(a small star and a low $ b $ can give the same transit duration as a large star and a high $ b $).
In the case of KIC~1255b,
it is still possible to derive constraints on $ a\sub{p} / R\sub{\star} $ from the light curve,
but, because the occulting object is a non-trivially shaped dust cloud rather than a solid sphere,
this requires a numerical approach.
Hence, we fit the phase-folded \textit{Kepler} light curve using
the \mbox{1-D} dust cloud model of \citet{2012A&A...545L...5B},
adding $ a\sub{p} / R\sub{\star} $ as an extra free parameter.\footnote{%
This model also contains implicit information about the stellar properties in the limb-darkening coefficient $ u $.
We tested running the model with $ u = 0.76 $ and $ u = 0.79 $,
appropriate for the two possible stellar types \citep{2000A&A...363.1081C},
and found the difference in outcome to be negligible.
Fig.~\ref{fig:a_vs_b} shows the results for the $ u = 0.79 $ run.
}
The outcome of the MCMC fitting shows a clear degeneracy between $ a\sub{p} / R\sub{\star} $ and $ b $ (see Fig.~\ref{fig:a_vs_b}).
Nevertheless, there is a firm ($3 \sigma$) constraint on the scaled semi-major axis of $ 2.8 \lesssim a\sub{p} / R\sub{\star} \lesssim 4.9 $,
which corresponds to a mean-stellar-density range of $ 0.9 \mathrm{~g~cm}^{-3} \lesssim \rho\sub{\star} \lesssim 5.4 \mathrm{~g~cm}^{-3} $.

\begin{figure}
  \includegraphics[width=\linewidth]{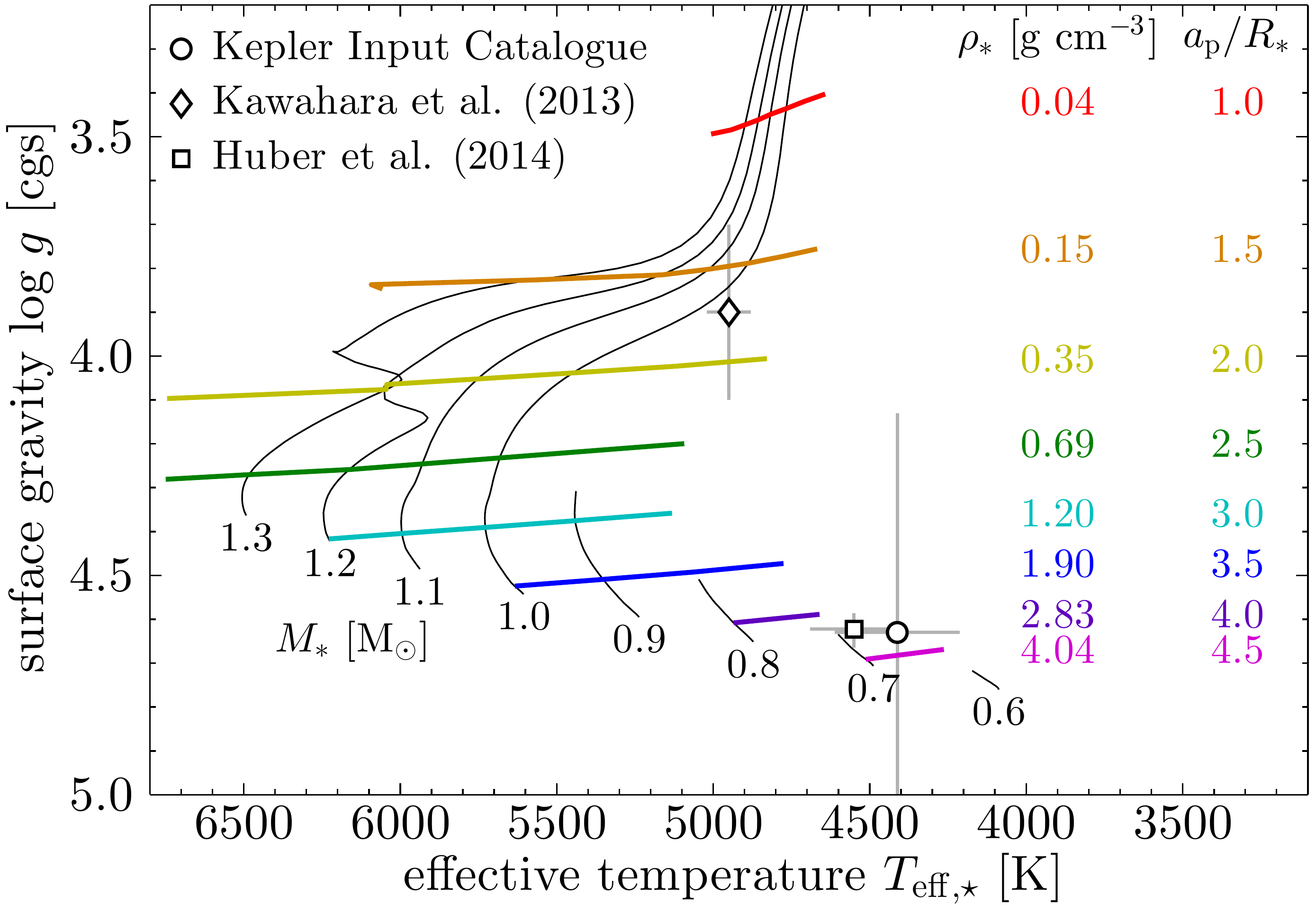}
  \caption{
  Estimates of the effective temperature~$ T_\mathrm{eff,\star} $ and surface gravity~$ \log \varg $
  of the star KIC~12557548,
  compared with predictions of stellar models.
  The symbols mark the different observational estimates of the stellar parameters from Table~\ref{tbl:kic_pars}.
  The thin black lines show Yonsei--Yale stellar evolutionary tracks
  for different stellar masses,
  indicated in solar units at the base of each track.
  The tracks start at the zero-age main sequence
  and they are truncated at an age of 13.8~Gyr
  (only relevant for stars with $ M_\star < 1.0 $~M$ _\odot $).
  As stars evolve, they move upward in this diagram, following the tracks toward lower values of $ \log \varg $.
  The thick coloured lines are curves of constant mean stellar density $ \rho\sub{\star} $.
  The table on the right shows the values of mean stellar density corresponding to the different lines
  and the values of the scaled semi-major axis $ a\sub{p} / R_\star $ of planet KIC~1255b that correspond to these densities.
  }
  \label{fig:isodens}
\end{figure}

To use this constraint to determine the evolutionary status of KIC~12557548,
we compare it with the mean stellar densities predicted for the two scenarios by stellar models.
Figure~\ref{fig:isodens} shows Yonsei--Yale stellar evolutionary tracks \citep{2003ApJS..144..259Y,2004ApJS..155..667D}
and the associated curves of constant mean stellar density.\footnote{%
The evolutionary tracks we show use
a metallicity of [Fe$/$H] = 0.05
and an $\upalpha$-enhancement of [$\upalpha/$Fe] = 0.0.
Varying the metallicity within the range suggested by Table~\ref{tbl:kic_pars}
does not change the conclusions.
}
It demonstrates that the stellar parameters found by \citet{2013ApJ...776L...6K}
correspond to subgiants with mean densities that are inconsistent with the constraints from the transit signal.
In contrast, the values of $ \rho\sub{\star} $ corresponding to the \mbox{mid-K} dwarf parameters
match well with the $ a\sub{p} / R\sub{\star} $ maximum in Fig.~\ref{fig:a_vs_b}.

Based on the stellar density constraint,
we reject the estimate of \citet{2013ApJ...776L...6K}
and conclude that KIC~12557548 is a \mbox{K-type} main-sequence star rather than a subgiant.
It is unclear why \citet{2013ApJ...776L...6K} arrive at a discrepant result
despite their more advanced method of determining stellar parameters
compared to the other efforts listed in Table~\ref{tbl:kic_pars}.
An intriguing possibility is that gasses released by the evaporating planet
contaminate the stellar spectrum, affecting the retrieval of the stellar parameters.
If this is the case, it gives hope to probing the composition of the planet more directly
by spectroscopically examining the gas component of the planet's outflow.

\section{Complex-refractive-index treatment}
\label{app:lnk_recipe}

\begin{figure}
  \includegraphics[width=\linewidth]{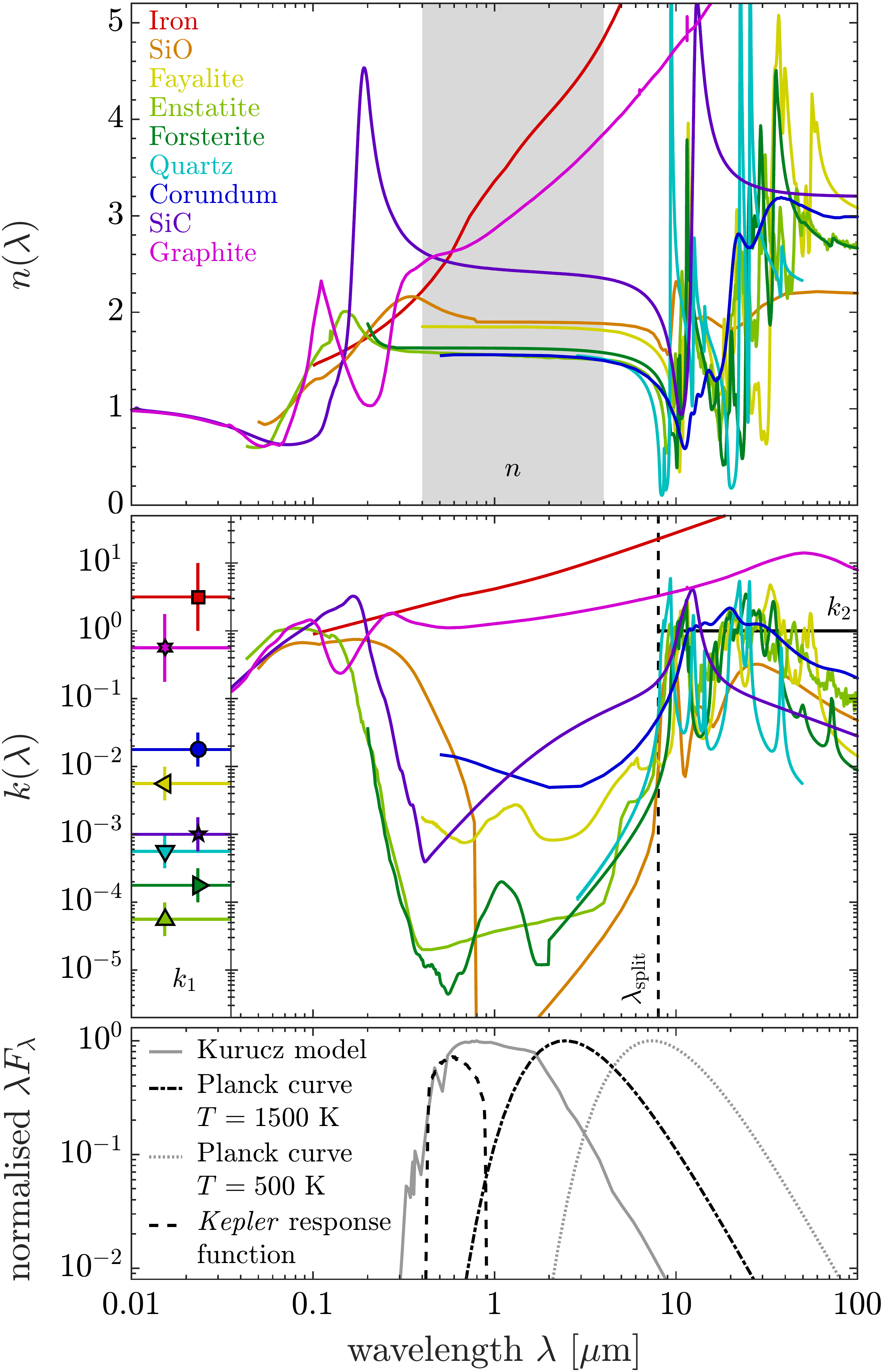}
  \caption{
  Complex refractive indices as a function of wavelength for the dust species considered in this study.
  \textbf{Top:}~The real part of the complex refractive index $ n ( \lambda ) $.
  The grey area indicates the range in wavelength used to determine the values of $ n $ listed in Table~\ref{tbl:dust_pars}.
  \textbf{Middle right:}~The imaginary part of the complex refractive index $ k ( \lambda ) $.
  The vertical dashed line at $ 8 \; \upmu$m marks $ \lambda\sub{split} $,
  which is the boundary between $ k_1 $ and $ k_2 $,
  the two constant values of the imaginary part of the complex refractive index used in our recipe.
  The horizontal black line at unity marks the value we use for $ k_2 $.
  \textbf{Middle left:}~The values of $ k_1 $ (listed in Table~\ref{tbl:dust_pars})
  that best reproduce the $ T\sub{d} ( s ) $ profiles of the various materials.
  The symbols are offset horizontally for visibility only.
  No symbol is shown for SiO, since its temperatures cannot be reproduced by our complex-refractive-index recipe.
  \textbf{Bottom:}~The spectrum of the Kurucz model atmosphere that we use for the host star KIC~12557548,
  together with thermal emission spectra of dust at different temperatures, all normalised to unity.
  Also shown is the \textit{Kepler} response function (not normalised),
  for which the vertical axis should read ``response''.
  }
  \label{fig:lnk}
\end{figure}

\begin{figure}
  \includegraphics[width=\linewidth]{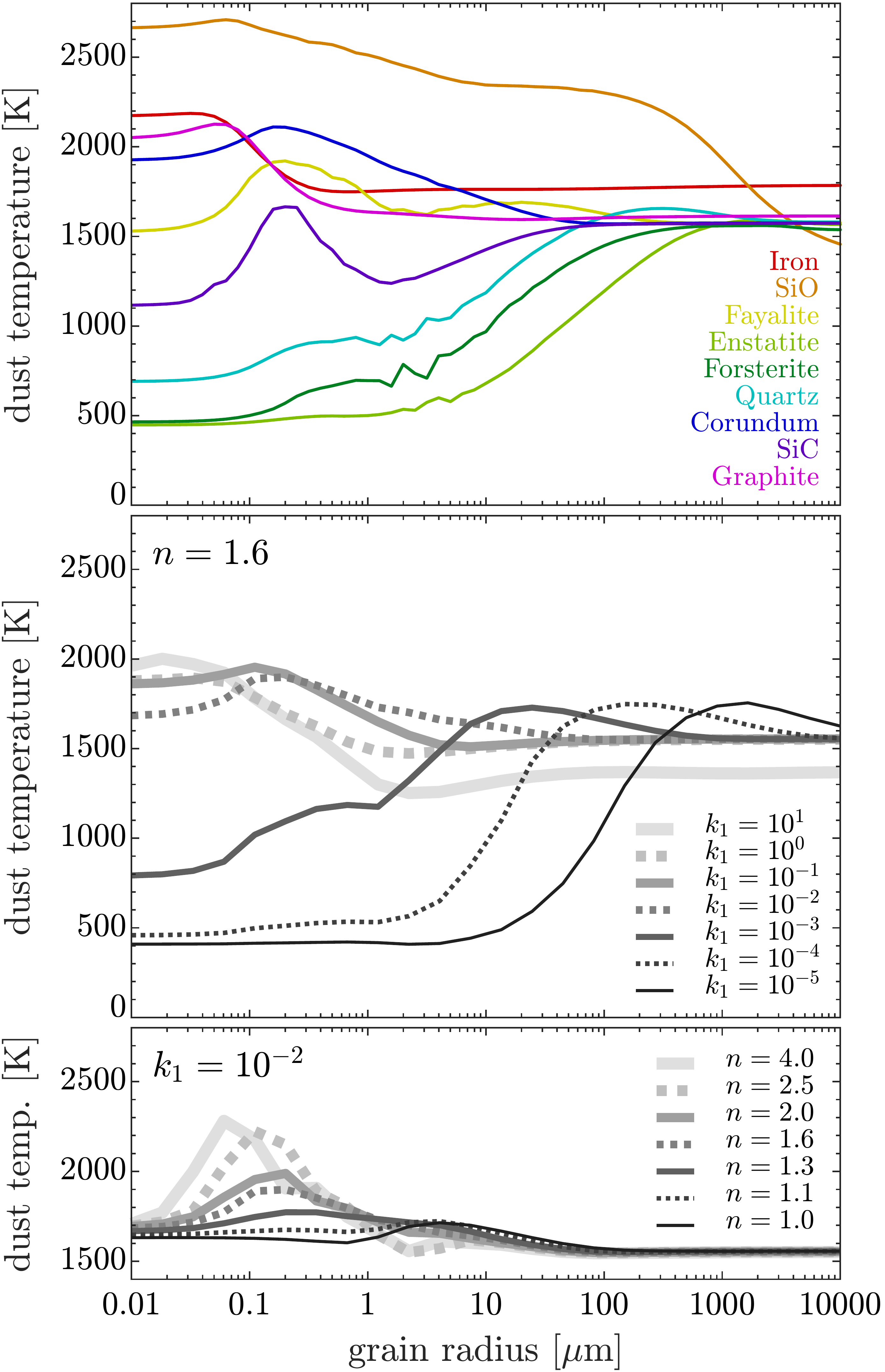}
  \caption{
  Dust temperatures at the distance of the planet as a function of grain size calculated using different complex refractive indices:
  \textbf{Top:}~Using the full wavelength-dependent complex refractive indices of real dust species, measured through laboratory experiments.
  \textbf{Middle:}~Using our simple complex-refractive-index prescription, keeping $ n $ fixed and varying $ k_1 $.
  \textbf{Bottom:}~Using our simple complex-refractive-index prescription, keeping $ k_1 $ fixed and varying $ n $.
  }
  \label{fig:temp_comp}
\end{figure}

Our dust-tail model uses a simple prescription for the complex refractive index of the dust material.
This recipe, summarised by Eq.~\eqref{eq:lnk_recipe}, consists of two free parameters: 
$ n $, the real part of the complex refractive index (one value for all wavelengths),
and $ k_1 $, the imaginary part of the complex refractive index at wavelengths below $ \lambda\sub{split} = 8 $~$\upmu$m.
For wavelengths of 8~$\upmu$m and higher, the imaginary part is fixed at $ k_2 = 1 $.
Our model uses the complex refractive index in Mie calculations to compute the optical efficiency factors of the dust grains.
These, in turn, are needed to find the $ \beta $ ratios and temperatures of the dust grains, and to generate the synthetic light curves.
In this appendix, we demonstrate that our simple recipe for the complex refractive index
can reproduce many of the dust temperatures found for real materials,
and we explain how we assign values of $ n $ and $ k_1 $ to the real materials as listed in Table~\ref{tbl:dust_pars}.

Figure~\ref{fig:lnk}
shows the full wavelength-dependent complex refractive indices of the nine materials considered in this study.
The sources of these data are listed in Table~\ref{tbl:dust_pars}.
For materials with different complex refractive indices for different crystal axes,
we combined the data corresponding to the different axes using the \citet{1935AnP...416..636B} mixing rule.
Also shown in Fig.~\ref{fig:lnk} are the \citet{1993KurCD..13.....K} model that we use for the stellar spectrum,
two Planck curves for possible dust temperatures, and the \textit{Kepler} response function \citep{2010ApJ...713L..79K}.
These can all be used to determine which parts of the $ n ( \lambda ) $ and $ k ( \lambda ) $ spectra
are the most important for the Mie calculations.

From Fig.~\ref{fig:lnk} it becomes clear that,
at visible and near-infrared wavelengths,
the values of $ k ( \lambda ) $ vary significantly between materials.
In the mid-infrared region, however, most materials surveyed here exhibit a range of narrow $ k ( \lambda ) $ features,
but have average $ k ( \lambda ) $ values that fall within a much smaller range.
This motivates our choice of a fixed $ k_2 = 1 $ for wavelengths beyond $ \lambda\sub{split} = 8 $~$\upmu$m.
The rise in $ k ( \lambda ) $ in the ultraviolet seen for many materials does not affect their heating,
since the stellar spectrum already drops off rapidly around $ \lambda = 0.3$~$\upmu$m.
This justifies the use of a constant $ k_1 $ for all wavelengths lower than 8~$\upmu$m.

Figure~\ref{fig:temp_comp} compares equilibrium dust temperatures calculated
using the full wavelength-dependent complex refractive indices
and the simple recipe of constant $ n $, $ k_1 $, and $ k_2 $.
By varying $ k_1 $, a large range of the temperatures observed for the real dust species can be reproduced.
Varying $ n $ has a smaller effect of the temperature profiles,
causing a bump in temperature at grain sizes of $ s \sim 0.01$~$\upmu$m for higher values of $ n $.

Of the materials we tested, the only one for which the simple recipe fails completely is silicon monoxide (SiO).
The reason for this is that the $ k ( \lambda ) $ profile of SiO changes abruptly around $ \lambda = 0.8$~$\upmu$m,
causing the grains to efficiently absorb the stellar radiation, but not efficiently cool in the near infrared.

The very low temperatures of some silicates (enstatite, forsterite, and quartz)
can be explained by their combination of low $  k ( \lambda ) $ in the optical and near infrared,
with the strong features in the mid infrared.
This causes them to absorb stellar radiation very inefficiently,
and only cool efficiently when they reach temperatures that are so low that the mid infrared features become relevant,
as illustrated by the 500~K Planck curve in Fig.~\ref{fig:lnk}.
Our complex-refractive-index recipe with $ \lambda\sub{split} = 8 $~$\upmu$m and $ k_2 = 1 $ manages to capture this behaviour well.

The complex refractive indices also affect other parts of the dust-tail model ($ \beta $ ratios and the generation of light curves),
but its effect on the final MCMC results through dust temperature is the most pronounced.
This is apparent from the fact that
$ n $ has a much smaller effect on dust temperatures than $ k_1 $ (see Fig.~\ref{fig:temp_comp})
and it is not well constrained by the data (see Sect.~\ref{s:res_constr}).

In order to compare the real dust species to the model results,
we wish to assign a single value of $ n $ and $ k_1 $ to each of the materials.
Figure~\ref{fig:lnk} shows that $ n( \lambda ) $ remains almost constant
in the wavelength range 0.4~$\upmu\mathrm{m} < \lambda < 4 $~$\upmu$m (grey area)
for all tested materials except iron and graphite, for both of which $ n( \lambda ) $ gradually rises with wavelength.
This wavelength regime includes the peak of the stellar spectrum, as well as the \textit{Kepler} bandpass.
To determine the value of $ n $ for each of the dust species, we therefore take the average value of $ n( \lambda ) $ over these wavelengths.
These values are listed in Table~\ref{tbl:dust_pars}.

The value of $ k ( \lambda ) $ of a given dust species can vary by orders of magnitude within the relevant wavelength regime.
Therefore, rather than trying to estimate an effective or average value directly from the $ k ( \lambda ) $ profile,
we take the value of $ k_1 $ for which our simple recipe yields the $ T\sub{d} ( s ) $ profile that best matches the one computed from the full data of that species.
To do this, we set $ n $ at the value
established before
and go through $ \log_{10} ( k_1 ) $ in steps of 0.25,
visually checking for a good match between the two $ T\sub{d} ( s ) $ profiles,
in particular at grain sizes of 0.01~$\upmu\mathrm{m} < s < 10 $~$\upmu$m.
The uncertainties in $ \log_{10} ( k_1 ) $ that we quote reflect the values beyond which the profiles begin to deviate significantly.

The results are listed in Table~\ref{tbl:dust_pars} and shown in the middle left panel of Fig.~\ref{fig:lnk}.
As mentioned before, our complex-refractive-index recipe fails to reproduce the $ T\sub{d} ( s ) $ profile of silicon monoxide.
For the remaining materials, the match is good, except for iron and corundum,
for which the temperatures predicted by the best-fitting $ n $ and $ k_1 $ values are slightly lower than the ones found using the full $ n( \lambda ) $ and $ k( \lambda ) $ data.
This discrepancy means that these materials will have sublimation rates that are somewhat higher than
suggested by
their position in \mbox{$ J\sub{1700\,K} $-$ k_1 $} space.

\end{appendix}

\end{document}